\documentstyle[12pt]{article}
\let\includefigures=\iftrue
\let\useamsfonts=\iftrue
\let\usefraktur=\iftrue
\includefigures
  \input epsf
  \def\figbox#1#2{\epsfxsize=#2\epsfbox{#1}}
\else
  \def\figbox#1#2{\strut\vspace{.5in}}
\fi
\useamsfonts
\font\blackboard=msbm10 scaled \magstep1
\font\blackboards=msbm7
\font\blackboardss=msbm5
\newfam\black
\textfont\black=\blackboard
\scriptfont\black=\blackboards
\scriptscriptfont\black=\blackboardss
\def\Bbb{\fam\black\relax}
\else
\def\Bbb{\bf}
\fi
\usefraktur
\font\eufmtn=eufm10 scaled \magstep1
\font\eufmsv=eufm7
\font\eufmfv=eufm5
\newfam\frakfam
\textfont\frakfam=\eufmtn
  \scriptfont\frakfam=\eufmsv
  \scriptscriptfont\frakfam=\eufmfv
\def\frak{\fam\frakfam\relax}
\font\bigfrak=eufm10 scaled \magstep2
\else
\def\frak{\bf}
\font\bigfrak=cmbx10 scaled \magstep2
\fi
\newcommand{\llangle}{\langle\hspace{-2pt}\langle}
\newcommand{\rrangle}{\rangle\hspace{-2pt}\rangle}
\newtheorem{definition}{Definition}[section]

\newtheorem{lemma}{Lemma}[section]
\newtheorem{proposition}{Proposition}[section]
\newtheorem{theorem}{Theorem}[section]
\newtheorem{corollary}{Corollary}[section]

\newcommand{\Int}{\mathop{\rm Int}\nolimits}
\newcommand{\ad}{\mathop{\rm ad}\nolimits}
\newcommand{\Ad}{\mathop{\rm Ad}\nolimits}
\newcommand{\Dom}{\mathop{\rm Dom}\nolimits}
\renewcommand{\Im}{\mathop{\rm Im}\nolimits}
\newcommand{\Isom}{\mathop{\rm Isom}\nolimits}
\newcommand{\Stab}{\mathop{\rm Stab}\nolimits}
\newcommand{\rank}{\mathop{\rm rank}\nolimits}
\newcommand{\Id}{\mathop{\rm Id}\nolimits}
\newcommand{\Exp}{\mathop{\rm Exp}\nolimits}
\newcommand{\Sym}{\mathop{\rm Sym}\nolimits}
\newcommand{\Hol}{\mathop{\rm Hol}\nolimits}
\newcommand{\Path}{\mathop{\rm Path}\nolimits}
\newcommand{\proj}{\mathop{\rm proj}\nolimits}
\newcommand{\pr}{\mathop{\rm pr}\nolimits}

\begin{document}
% set footnotes to be symbols
\def\thefootnote{\fnsymbol{footnote}}
%title
\begin{titlepage}
\noindent
March~1995\hfill{UMTG--184}\newline
\strut\hfill{\tt hep-th/9503226}
\par\vskip 2cm
\begin{center}
{\large \bf Target-Space Duality between}\\
{\large \bf Simple Compact Lie Groups and Lie Algebras}\\
{\large\bf under the Hamiltonian Formalism:}\\
{\large\bf I. Remnants of Duality at the Classical Level}\\[1cm]
%author
{\large Orlando Alvarez%
\footnote{\footnotesize\tt e-mail: alvarez@phyvax.ir.miami.edu}
 \hspace{2.5cm}
    Chien-Hao Liu%
\footnote{\footnotesize\tt e-mail: chienliu@phyvax.ir.miami.edu}}\\[0.5cm]
%address
{\it Department of Physics}\\
{\it University of Miami}\\
{\it P.O. Box 248046}\\
{\it Coral Gables, FL 33124}\\[1cm]
\end{center}

%abstract%
\begin{abstract}
It has been suggested that a possible classical remnant of the phenomenon
of target-space duality (T-duality) would be the equivalence of the
classical string Hamiltonian systems.  Given a simple compact Lie group $G$
with a bi-invariant metric and a generating function $\Gamma$ suggested in
the physics literature, we follow the above line of thought and work out
the canonical transformation $\Phi$ generated by $\Gamma$ together with an
$\Ad$-invariant metric and a B-field on the associated Lie algebra $\frak
g$ of $G$ so that $G$ and $\frak g$ form a string target-space dual pair at
the classical level under the Hamiltonian formalism.  In this article, some
general features of this Hamiltonian setting are discussed.  We study
properties of the canonical transformation $\Phi$ including a careful
analysis of its domain and image.  The geometry of the T-dual structure on
$\frak g$ is lightly touched.  We leave the task of tracing back the
Hamiltonian formalism at the quantum level to the sequel of this paper.
\end{abstract}

%end topmatter%
\vspace{0.1cm}

\noindent
{\bf Acknowledgments} We would like to thank Hung-Wen
Chang for very helpful discussions, Thom Curtright for sharing with
us his joint work with C. Zachos, and Marco Monti for help with Xfig.

\end{titlepage}

% Body of paper
% reset foonotes and section counter
\def\thefootnote{\arabic{footnote}} \setcounter{footnote}{0}
\setcounter{section}{-1}
\section{Introduction and Outline}
\subsection{Introduction}

Target space duality (T-duality) is a very surprising phenomenon in string
theory\footnote{See the review \cite{G-P-R} for a comprehensive set of
references.}.  In essence, two target-spaces are dual to each other if both
lead to the same string theory.  The usual technical definition involves
using path-integrals to sum over the space of all smooth maps from surfaces
(string world-sheets) to target manifolds~\cite{B1,B2,F-J,R-V,G-R1,M-V}.  In
this aspect, it is a quantum mechanical phenomenon.  Nevertheless, it is
natural to ask:
\par\noindent\strut
\hspace{.75in} {\it Q:}\quad\parbox[t]{4in}{\it Are
there classical aspects of the phenomenon of target space duality?\/}
\par\vskip 8pt\noindent
As already pointed out in the literature (e.g.
\cite{A-AG-B-L,A-AG-L2,C-Z,G-P-R,G-R3,G-R-V}),
one possible answer may be the equivalence of the associated string
Hamiltonian systems.

As will be discussed in Sec.~1.3, for the simplest known example, the
$(R\leftrightarrow\frac{1}{R})$-duality for $S^1$, the above naive picture
after appropriate modification captures many features of target-space
duality.  Backed by this example and some lessons learned from it, we next
turn our attention to another known example in the physics literature
\cite{A-AG-B-L,A-AG-L1,C-Z,dlO-Q,E-G-R-S-V,G-K,G-R2,G-R-V}.
Recall that the simple compact Lie group
$SU(2)$ and its associated Lie algebra form a target-space dual pair when
$SU(2)$ is endowed with a bi-invariant metric and its associated Lie
algebra ${\frak su}(2)$ is endowed with a metric and a $B$-field which, up
to a constant multiple, are written in linear coordinates respectively as
\cite{C-Z}
$$
\tilde{g}_{ij}\;=\;\frac{\left(\delta_{ij}+4v^i v^j\right)}{1+4v^2}
\mbox{\hspace{0.5cm} and \hspace{0.5cm}}
\widetilde{B}_{ij}\;=\;\frac{\epsilon_{ijk}v^k}{1+4v^2}.
$$
In terms of Hamiltonian systems, the duality of this pair comes from
a formal canonical transformation from the loop space $LT^{\ast}SU(2)$
to the loop space $LT^{\ast}{\frak su}(2)$.
This canonical transformation is generated by
$$
\Gamma(g,v)\;=\;\int_{S^1}\,Tr\left(v\,g^{-1}dg\right)
$$
in coordinate-free, fundamental matrix form.
The latter expression is immediately applicable for general
Lie groups and their associated Lie algebras.
This observation leads us to this present work.

Recently, a geometrical picture of duality~\cite{K-S1,K-S2} has emerged
which allows one to write down the general duality transformation when
there is a group action on a manifold. In the present paper, we use the
formalism of \cite{C-Z} to look more closely at the example of the target
being a simple compact Lie group.

In brief, given a simple compact Lie group $G$ with a bi-invariant metric,
let $\frak g$ be its associated Lie algebra.  We take the generating
function $\Gamma$ as the foundation of our approach and work out the
canonical transformation $\Phi$ it generates from $LT^{\ast}G$ to
$LT^{\ast}{\frak g}$.  We obtain also an $\Ad$-invariant metric and an
$\Ad$-invariant $B$-field (a 2-form) on the associated Lie algebra $\frak
g$ so that they form a T-dual pair at the classical level under the
Hamiltonian formalism.  This could possibly be an exact dual pair in terms
of path-integrals at the quantum level.  In this first paper, we focus on
properties of the canonical transformation $\Phi$ and the T-dual geometry
on $\frak g$ and leave the important issue of how exactly $G$ and $\frak g$
form a dual pair at the quantum level to the sequel.

Recently there appeared an article \cite{Lo} by Y. Lozano on the same
subject. Interested readers may compare our setting here with hers.

\vspace{1cm}

\subsection{Outline}
\begin{quote}
  {\bf 1. Target-Space Duality in Hamiltonian Formalism}
    \begin{quote}
      1.1 Hamiltonian formalism for string theory.

      1.2 Target-space duality.

      1.3 A lesson from the
           $(R\leftrightarrow\frac{1}{R})$-duality of $S^1$.
    \end{quote}

  {\bf 2. The T-Dual Transformation and T-Dual Structures}
    \begin{quote}
      \begin{tabbing}
      2.1 \= A generating function and the induced canonical trans-\\
          \> formation.
      \end{tabbing}

      2.2 The dual structures on the associated Lie algebra.

      2.3 A second glance at $\Phi$.

      2.4 Symmetries in the theory.

    \end{quote}

  {\bf 3. The Geometry of $({\frak g},\llangle\:,\:\rrangle\,B)$}
    \begin{quote}
      3.1 Preliminaries to study the T-dual geometry.

      3.2 An $\Ad$-invariant polarization in ${\frak g}-V_0$.

      3.3 Basic properties of the T-dual geometry.

      3.4 Riemannian geometry of the T-dual metric.

      3.5 The B-field $B$.

    \end{quote}
\end{quote}

\newpage

\section{Target-Space Duality in Hamiltonian Formalism}

\subsection{Hamiltonian formalism for string theory}

In string theory a particle is assumed to be a one-dimensional extended
object.  There are two kinds of them, open and closed strings.  In this
article we shall restrict ourselves only to closed strings, given by smooth
maps from $S^1$ into a smooth target manifold.

Neglecting the dilaton and other fields, the target-space data for a string
theory consists of a Riemannian manifold with a 2-form (usually called a
$B$-field by physicists) $(M,ds^2,B)$.  We shall denote it collectively by
$M$ when both the Riemannian metric and the $B$-field are understood from
the text.  The configuration space consists of all possible positions of
the particle and hence is given by the loop space
$$
LM\;=\;\{\phi:S^1\longrightarrow M\,|\,\phi \mbox{ is $C^{\infty}$.}\}.
$$
The phase space requires however some choices. Since we are only
interested in objects describable as smooth objects along a circle,
we choose the phase space to be $LT^{\ast}M$ instead of the much larger
$T^{\ast}LM$. There is  a canonical symplectic structure
{\boldmath $\omega$} on $LT^{\ast}M$ induced from
the canonical symplectic structure $\omega$ on $T^{\ast}M$ given by
$$
\mbox{\boldmath $\omega$}_{\gamma}(\eta,\xi)\;=\;
   \int_{S^1}\:d\sigma\,
       \omega\left(\eta_{\gamma(\sigma)},\xi_{\gamma(\sigma)}\right),
$$
where $\gamma$ is in $LT^{\ast}M$ and $\eta$, $\xi$ are two tangent
vectors at $\gamma$. They are simply two vectors fields in $T^{\ast}M$
along the loop $\gamma$.

The Lagrangian density from the (1+1)-dimensional $\sigma$-model over a
cylinder can be thought of as an energy function for paths in the
configuration space.  It can be rephrased as a Lagrangian $\cal L$ defined
on the tangent bundle
$T_{\ast}LM\,=\,LT_{\ast}M$ of $LM$. Denote a point in $T_{\ast}LM$ by
$(\phi,X)$ where $\phi\,\in\,LM$ and $X$ is a smooth vector field in $M$
along $\phi$, then the Lagrangian can be written as
$$
{\cal L}(\phi,X)\;=\;\int_{S^1}\,d\sigma\,{\cal L}(\phi,X;\sigma)
$$
with
$$
{\cal L}(\phi,X;\sigma)\;=\;
  \frac{1}{2}\left(\langle X(\sigma),X(\sigma)\rangle\,-\,
   \langle\phi_{\ast}\partial_{\sigma},
       \phi_{\ast}\partial_{\sigma}\rangle\right)\:+\:
           B\left(X(\sigma),\phi_{\ast}\partial_{\sigma}\right),
$$
where $\langle\;,\;\rangle$ stands for the metric on $M$, and
$\partial_{\sigma}$ is the coordinate vector field along
$S^1$.

The canonical momentum density $\pi$ associated to $\cal L$ for
$(\phi,X)$ is given by
$$
\pi(\sigma)\;=\;\frac{\delta{\cal L}}{\delta X}(\sigma)\;=\;
    \langle\;\cdot\;,X(\sigma)\rangle\:+\:
       B\left(\;\cdot\;,\phi_{\ast}\partial_{\sigma}\right).
$$
The Legendre transformation now takes functions on $LT_{\ast}M$ to
functions on $LT^{\ast}M$. The image of the Lagrangian $\cal L$ becomes
the string Hamiltonian function ${\cal H}$ on $LT^{\ast}M$. Its
density function along $S^1$ is given by
$$
{\cal H}(\phi,\pi;\sigma)\;=\;\frac{1}{2}
 \langle\pi(\sigma)-B\left(\;\cdot\;,\phi_{\ast}\partial_{\sigma}\right)
   \:,\:\pi(\sigma)-
     B\left(\;\cdot\;,\phi_{\ast}\partial_{\sigma}\right)\rangle^{\sim}
        \;+\;\frac{1}{2}\langle\phi_{\ast}\partial_{\sigma}\:,\:
                      \phi_{\ast}\partial_{\sigma}\rangle,
$$
where $\langle\;,\;\rangle^{\sim}$ stands for the induced metric on the
fiber of $T^{\ast}M$ from $\langle\;,\;\rangle$.

Basically all the information about the classical
physics for a closed string is contained in this Hamiltonian system.

\subsection{Target-space duality}

Though the term ``{\it target-space duality}'' has become more or less
official in the literature, a better name for it would be ``{\it
string-equivalence between target-spaces}'' \cite{A-G-M}.  The latter says
exactly the meaning hidden under the former.  Technically, this means that
there exists a correspondence $\Phi$ that takes the states and observables
in the string theory associated to one target-space $(M, ds^2, B, \cdots)$
to those of the string theory associated to another target-space
$(\widetilde{M}, \widetilde{ds^2}, \widetilde{B}, \cdots)$ such that the
related correlation functions are all identical.  Thus, as long as physics
is concerned, one cannot tell whether the particle is moving about in one
or another target-space in the same equivalence class.  Since these
correlation functions are all defined formally via Feynman's path-integral,
the definition indicates that target-space duality is actually a quantum
level phenomenon.  One would like to know {\it if this phenomenon manifests
itself at the classical level}.

Since all the information of the classical physics for a string theory is
completely contained in the string Hamiltonian system described in Sec.~1.1,
a naive guess for the classical remnant of target-space duality is the
equivalence of string Hamiltonian systems.  This equivalence would be given
by a canonical transformations between string phase spaces that take the
string Hamiltonian function on one phase space to that on another.  In the
next section we shall do a redemonstration of a known example where the
target space is a circle.  The classical remnant of target space duality
with be a classical Hamiltonian equivalence except that the respective
phase spaces have to be restricted.  In this example ``classical duality''
only exists between {\it reduced Hamiltonian systems\/}.  This may be a
general feature of target space duality.  Namely, its classical remnant is
a reduced Hamiltonian system though a general rule for this classical
reduction is not known yet.

In this paper we will explore these issues for the case of a $G-{\frak
g}$ pair.

\subsection{ A lesson from the $(R\leftrightarrow\frac{1}{R})$-duality of
$S^1$}

The $S^1$-target case is the simplest and best known example of
target-space duality.  It indicates a new relation between physics in the
small scale and physics in the large scale.  Such a relationship may be
applicable to the removal of initial singularities of a space-time in
general relativity.

The target-space in this example is $S^1_R$, a circle of radius $R$.
The phase space is $LT^{\ast}S^1_R$. Let $(\theta,\pi)$ be a canonical
coordinate system for $T^{\ast}S^1_R$,  where $\theta$ runs over the
interval $[0,2\pi]$ and is proportional to the length of the circle.
Let $\gamma(\sigma)\,=\,(\theta(\sigma),\pi(\sigma))$ be a loop in
$T^{\ast}S^1_R$. Then the value of the Hamiltonian function
at $\gamma$ is given by
$$
{\cal H}(\gamma)\;=\;\int_{S^1}\,d\sigma\,{\cal H}(\gamma;\sigma)
$$
with
$$
{\cal H}(\gamma;\sigma)\;=\;\frac{1}{2R^2}\pi(\sigma)^2\:+\:
   \frac{R^2}{2}\left(\frac{d\theta}{d\sigma}\right)^2.
$$
The Hamiltonian vector field $X_{\cal H}$
on $LT^{\ast}S^1_R$ associated to $\cal H$ can be computed
straightforwardly. Its value at a $\gamma$ in $LT^{\ast}S^1_R$
is a vector field along $\gamma$ in $T^{\ast}S^1_R$ and is given by
$$
X_{\cal H}|_{\gamma}(\sigma)\;=\;
\left(R^2\frac{d^2\theta}{d\sigma^2}\right)\,\partial_{\pi}|_{\gamma(\sigma)}
\:+\:\left(\frac{1}{R^2}\pi(\sigma)\right)\,
                                    \partial_{\theta}|_{\gamma(\sigma)}.
$$

The form of the Hamiltonian function suggests immediately
a transformation $\Phi$ from $LT^{\ast}S^1_R$ to
$LT^{\ast}S^1_{\frac{1}{R}}$ that leaves the form invariant:
$$
\Phi(\gamma)(\sigma)\;=\;
    \left(\int_0^{\sigma}\:d\varsigma\:\pi(\varsigma)\;,\;
        \frac{d\theta}{d\sigma}(\sigma)\right).
$$
Let $\tilde{\gamma}\:=\:\Phi(\gamma)$, then $\widetilde{\cal H}$,
the pushed-forward of $\cal H$ to $LT^{\ast}S^1_{\frac{1}{R}}$, has density
$$
\widetilde{\cal H}(\tilde{\gamma};\sigma)\;=\;
  \frac{1}{2R^2}\left(\frac{d\tilde{\theta}}{d\sigma}\right)^2
   \:+\:\frac{R^2}{2}\tilde{\pi}^2,
$$
which is exactly the string Hamiltonian function with target space
$S^1_{\frac{1}{R}}$. Unfortunately, as we shall see,
this natural candidate for the sought
canonical transformation is not extendable to the whole phase
space. Nevertheless, a natural ``{\it quantization condition}''
comes in to select the correct reduced phase space on which
everything works.

\begin{figure}
 \centerline{\figbox{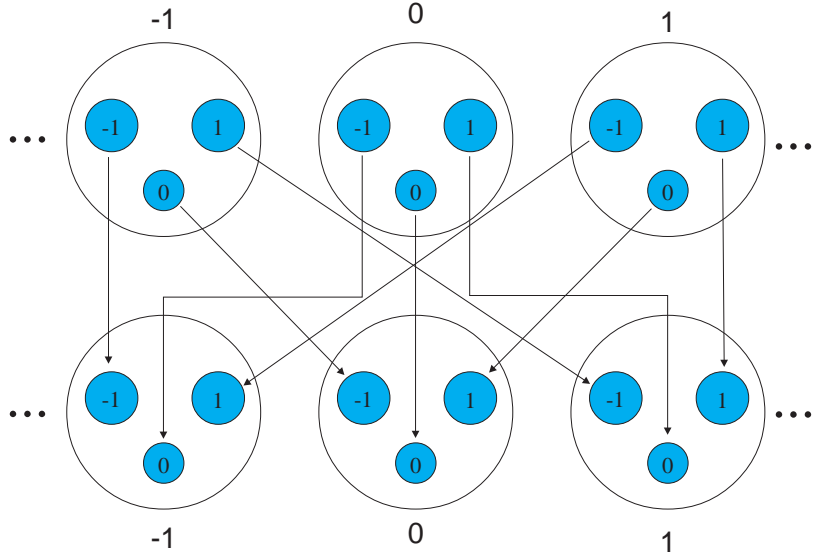}{11cm}}
 \caption[circle-dual]{T-duality between $S^1_R$
            and $S^1_{\frac{1}{R}}$. The outer labels denote the
            winding ``quantum'' number; the inner labels denote the
            momentum ``quantum'' number.}
  \protect\label{circle-dual}
\end{figure}

\begin{proposition}
Let $\gamma_{\theta}$ and $\gamma_{\pi}$ be respectively the $\theta$-
and $\pi$-component of a loop $\gamma$ in $T^{\ast}S^1_R$. Let
$$
L_R^{(m,n)}\;=\;\{\gamma\,|\,deg\,\gamma_{\theta}\,=\,m,\:
      \int_{S^1}\,d\sigma\,\pi(\sigma)\,=\,2\pi n\,\}.
$$
Then
\begin{enumerate}
  \item each $L_R^{(m,n)}$ is a sub-Hamiltonian system in
        $\left(LT^{\ast}S^1_R,\mbox{\boldmath $\omega$},\cal H\right)$;
  \item $\Phi$ is a canonical transformation from
        $L_R^{(m,n)}$ onto $L_{\frac{1}{R}}^{(n,m)}$.
\end{enumerate}
Schematic representation of $\Phi$ may is depicted in
{\sc (Figure~\ref{circle-dual})}.
\end{proposition}

\noindent{\it Proof:} Let $t$ be the parameter for the string
Hamiltonian flow. Continuity of the flow implies that winding number
of $\gamma_{\theta}$ is invariant. Using the explicit expression for
$X_{\cal H}$, the first claim then follows from the fact that
along the flow
\begin{eqnarray*}
\lefteqn{ \frac{d}{dt}\int_{S^1}d\sigma\,\pi(\sigma,t)\;=\;
     \int_{S^1}d\sigma\,\frac{\partial}{\partial t}\pi(\sigma,t)}\\
  & & =\;R^2\,\int_{S^1}d\sigma\,\frac{\partial^2}{\partial\sigma^2}
        \theta(\sigma,t)\;
      =\;R^2\,\int_{S^1}d\sigma\,\frac{\partial}{\partial\sigma}
        \left(\frac{\partial}{\partial\sigma}\theta(\sigma,t)\right),
\end{eqnarray*}
which vanishes since $T_{\ast}S^1$ is trivial and, thus,
$\frac{\partial}{\partial\sigma}\theta(\sigma,t)$ can be regarded
as a map from $S^1$ to $\Bbb R$.

That $\Phi$ maps $L_R^{(m,n)}$ onto $L_{\frac{1}{R}}^{(n,m)}$ is clear.
Its inverse is given by
$$
\Phi^{-1}(\tilde{\gamma})(\sigma)\;
  =\;\left(\int_0^\sigma ds\,\tilde{\pi},
                \frac{d}{d\sigma}\tilde{\theta}(\sigma)\right).
$$
One can check that $\Phi^{\ast}\widetilde{\cal H}\,=\,{\cal H}$.

Now let $Y$ be in $T_{\gamma}L^{(m,n)}_R$. In the canonical coordinates,
as a vector field along $\gamma$ in $T^{\ast}S^1$, one may write
$$
Y(\sigma)\;=\;A(\sigma)
  \left.\frac{\partial}{\partial\theta}\right|_{\gamma(\sigma)}\,
   +\,B(\sigma)
  \left.\frac{\partial}{\partial\pi}\right|_{\gamma(\sigma)}.
$$
Then
$$
\begin{array}{ccc}
 \Phi_{\ast} \; : \; T_{\gamma}L^{(m,n)}_R & \longrightarrow
          & \hspace{-1cm} T_{\Phi(\gamma)}L^{(n.m)}_{\frac{1}{R}}\\
   \displaystyle A(\sigma)
     \left.\frac{\partial}{\partial\theta}\right|_{\gamma(\sigma)}\,
       +\,B(\sigma)
           \left.\frac{\partial}{\partial\pi}\right|_{\gamma(\sigma)}
    & \longmapsto
    & \displaystyle
    \left(\int_0^{\sigma}ds\,B(s)\right)
       \left.\frac{\partial}{\partial\tilde{\theta}}
              \right|_{\Phi(\gamma)(\sigma)}\,
        +\,\left(\frac{d}{d\sigma}A(\sigma)\right)
            \left.\frac{\partial}{\partial\tilde{\pi}}
              \right|_{\Phi(\gamma)(\sigma)}\,.
\end{array}
$$
{}From this one can check straightforwardly that
$$
\tilde{\mbox{\boldmath $\omega$}}(\Phi_{\ast}Y_1,\Phi_{\ast}Y_2)\;
   =\;\mbox{\boldmath ${\omega}$}(Y_1,Y_2);
$$
and hence $\Phi$ is a symplectomorphism from $L^{(m,n)}_R$ onto
$L^{(n.m)}_{\frac{1}{R}}$. This concludes the proof.

\hspace{12cm}$\Box$
\medskip

\section{The T-Duality Transformation and T-Dual Structures}

With the preparation in Sec.~1 we shall focus for the rest of this
article on the case of simple compact Lie groups and
their associated Lie algebras. To avoid
confusion with other duals in the discussion, we will write ``T-dual''
for ``target-space dual''.

\subsection{A generating function and the induced canonical transformation}

\subsubsection{A natural generating function
$\Gamma:  L{\frak g}\times LG\longrightarrow \,{\Bbb R}$.}

Let $G$ be a simple compact Lie group and $\frak g$ be its associated
Lie algebra. We shall identify $\frak g$ constantly with $T_eG$, the
tangent space of $G$ at the identity $e$ or occasionally with the
space of all left-invariant vector fields on $G$ whenever necessary.
$G$ admits a bi-invariant positive-definite
metric which is unique up to a constant multiple.
This metric is proportional
to the Killing form of $G$. Its restriction to ${\frak g} = T_eG$
provides an $\Ad$-invariant inner product in the Lie algebra.
For simplicity of notation, we shall denote both of them by
$\langle\:,\:\rangle$.

Let $\Omega$ be the left invariant Maurer-Cartan 1-form of $G$.
Recall that, for $X\in T_gG$, it is defined by
$$
\Omega(X)=(l_{g^{-1}})_{\ast}(X)\in T_eG={\frak g},
$$
where $l_g:G\to G$ is left multiplication by $g$.

The we choose a generating function $\Gamma$ defined as follows.

\begin{definition}[Generating function]
  Let $(\psi: S^1\longrightarrow{\frak g})\in L{\frak g}$ and
  \newline $(\varphi: S^1\longrightarrow G)\in LG.$
  With $S^1$ parameterized by $\sigma$, we define
  $$
   \Gamma\left(\psi,\varphi;\sigma\right)=\langle\psi\left(\sigma\right),
   \Omega\left(\varphi_{\ast}\partial_{\sigma}\right)\rangle,
  $$
  where $\partial_{\sigma}$ is the coordinate vector field along
  $S^1$. We choose the generating function
  $\Gamma\,:\,L{\frak g}\times LG\,\longrightarrow\,{\Bbb R}$ to be
  $$
  \Gamma(\psi,\varphi) = \int_{S^1}d\sigma\Gamma(\psi,\varphi;\sigma).
  $$
\end{definition}

\noindent
{\it Remark:} Notice that when $G$ is identified with
a classical matrix group, the above expression for $\Gamma$ is exactly
$$
\Gamma(\psi,\varphi)= \mbox{\rm constant\/}\cdot
\int_{S^1}Tr\left(\psi\left(\sigma\right)\varphi\left(\sigma\right)^{-1}
\frac{d}{d\sigma}\varphi\left(\sigma\right)\right),
$$
which appears already in the literature for constructing dual models of
the chiral $SU(2)$-model.

\subsubsection{The induced canonical transformations}

In the following arguments we shall denote points in
$LT^{\ast}G$ by $(\varphi,\varpi)$ where $\varphi$ is a smooth
map from $S^1$ into $G$ and $\varpi$ is a 1-form along $\varphi$.
Similarly we shall denote points in $LT^{\ast}{\frak g}$
by $(\psi, \pi)$ where $\psi$ is a smooth map from
$S^1$ into ${\frak g}$ and $\pi$ is a 1-form along $\psi$.
Our first task is to work out the functional derivative
$$
\varpi=\frac{\delta\Gamma(\psi,\varphi)}{\delta\varphi},\hspace{1cm}
\pi=-\frac{\delta\Gamma(\psi,\varphi)}{\delta\psi};
$$
and then to solve $(\psi,\pi)$ and $(\varphi,\varpi)$ in terms of each
other.

\begin{proposition} The functional derivatives of the generating
function $\Gamma$ with respect to its arguments are respectively
\begin{eqnarray*}
  \varpi\; = &\frac{\delta\Gamma(\psi,\varphi)}{\delta\varphi}
         &= -\langle\left(\frac{d}{d\sigma}+
   \ad_{\Omega\left(\varphi_{\ast}\partial_{\sigma}\right)}
   \right)\psi,
   \Omega\left(\cdot\right)\:\rangle,\\
  \pi\; = &-\frac{\delta\Gamma(\psi,\varphi)}{\delta\psi}
    &= -\langle\Omega\left(\varphi_{\ast}\partial_{\sigma}\right),
   \cdot\:\rangle,
\end{eqnarray*}
where $\ad_{(\cdot)}$ is the $\ad$-representation of $(\cdot)$ on ${\frak g}$.
\end{proposition}

\medskip

\noindent
{\it Remark:} One may notice that, in the expression,
the part,
$$\langle
\ad_{\Omega(\varphi_{\ast}\partial_{\sigma})}\psi,\Omega(\,\cdot\,)\rangle,
$$
up to a constant, is exactly the canonical 3-form $\Xi$ on a simple Lie
group defined by
$$
\Xi(X,Y,Z)=K\left([\Omega(X),\Omega(Y)],\Omega(Z)\right),
$$
where $X$, $Y$, $Z$ are some tangent vectors at some point in $G$
, $[\;,\;]$ is the Lie bracket for the associated Lie algebra ${\frak g}$
and $K$ is the Killing form of $G$.

\medskip

\noindent{\it Proof:} Since $\Gamma(\psi,\varphi)$ is linear
with respect to $\psi$, one has immediately
$$
  -\frac{\delta\Gamma(\psi,\varphi)}{\delta\psi}=
    -\langle\Omega(\varphi_{\ast}\partial_{\sigma}),\:\cdot\:\rangle.
$$
Thus for the rest of the proof we shall focus on the computation of
$\frac{\delta\Gamma(\psi,\varphi)}{\delta\varphi}$.

\medskip

\noindent({\it i}) Let $X$ be a vector field along
$\varphi$. Let
\begin{eqnarray*}
\Upsilon\,:&S^1\times (-\varepsilon,\varepsilon)&\longrightarrow G\\
     &(\sigma,\tau)&
\end{eqnarray*}
such that
$$
\Upsilon(\cdot,0)\,=\,\varphi,
       \hspace{1cm}\Upsilon_{\ast}(\partial_{\tau}|_{\tau=0})\,=\,X.
$$
Let $\varphi_{\tau}\,=\,\Upsilon(\:\cdot\:,\tau)$. One has
\begin{eqnarray*}
\lefteqn{\int_{S^1}\frac{\delta\Gamma(\psi,\varphi)}{\delta\varphi}(\sigma)
    \left(X(\sigma)\right)d\sigma}\\
  & & =\; \left.\frac{d}{d\tau}\right|_{\tau=0}\Gamma(\psi,\varphi_{\tau})\\
  & & =\; \int_{S^1}\left.\frac{\partial}{\partial\tau}\right|_{\tau=0}
            \langle\psi(\sigma),
             \Omega\left(\varphi_{\tau\ast}\partial_{\sigma}
             \right)\rangle d\sigma\\
  & & =\; \int_{S^1}\langle\psi(\sigma),
               \left.\frac{\partial}{\partial\tau}\right|_{\tau=0}
       \Omega\left(\varphi_{\tau\ast}\partial_{\sigma}\right)\rangle d\sigma.
\end{eqnarray*}

\medskip

\noindent({\it ii}) We shall show next that
$$
\left.\frac{\partial}{\partial\tau}\right|_{\tau=0}
       \Omega\left(\varphi_{\tau\ast}\partial_{\sigma}\right)\;
   =\; \frac{d}{d\sigma}\Omega(X)
     + \ad_{\Omega\left(\varphi_{\ast}\partial_{\sigma}\right)}\Omega(X).
$$

Let $S=\Upsilon_{\ast}\partial_{\sigma},\:
                        T=\Upsilon_{\ast}\partial_{\tau}$.
Let $e_i$ be a basis for $\frak g$ and $\omega^i$ be
the 1-forms on $G$ obtained by left-translating the dual basis of $e_i$
around $G$. Then $\Omega=e_i\otimes\omega^i$; and
\begin{eqnarray*}
\lefteqn{\left.\frac{\partial}{\partial\tau}\right|_{\tau=0}
   \Omega\left(\varphi_{\tau\ast}\partial_{\sigma}\right)}\\
  & & =\; X\left(e_i\otimes\omega^i(S)\right)\;
         =\;e_i\cdot T\omega^i(S)|_{\tau=0}\\
  & & =\; e_i\cdot\{S\omega^i(T)+\omega\left([T,S]\right)+
         2\,d\omega^i(T,S)\}_{\tau=0}\\
  & & =\; S\Omega(X)+2\,d\Omega(T,S),
        \hbox{ since }T_{\tau=0}=X \hbox{ and } [T,S]=0;\\
  & & =\; \frac{d}{d\sigma}\Omega(X)\,+\,2\,d\Omega(T,S).
\end{eqnarray*}
By the Maurer-Cartan equation, i.e.
$$
d\Omega+\frac{1}{2}[\Omega,\Omega]=0,
$$
where $[\:,\:]$ means Lie bracket for the Lie algebra part and wedge product
for the 1-form part,
the second term in the last can be rewritten as
\begin{eqnarray*}
\lefteqn{  2\,d\Omega(T,S)\; =\; -[\Omega,\Omega](T,S)  }\\
  & & =\; [\Omega(S),\Omega(T)]\;
    =\; \ad_{\Omega\left(\varphi_{\ast}\partial_{\sigma}\right)}\Omega(X),
          \hbox{ at }\tau=0.
\end{eqnarray*}

\medskip

\noindent({\it iii}) Finally
from the fact that $\ad$ is skew-symmetric with respect to
$\langle\:,\:\rangle$, we have
\begin{eqnarray*}
 \int_{S^1}d\sigma\,\langle\psi(\sigma),\frac{d}{d\sigma}\Omega(X)
      + \ad_{\Omega\left(\varphi_{\ast}\partial_{\sigma}\right)}
         \Omega(X)\rangle
   & = &\int_{S^1}d\sigma\,\frac{d}{d\sigma}\langle\psi(\sigma),
       \Omega(X)\rangle\\
       && \mbox{}-
       \int_{S^1}d\sigma\,\langle\frac{d}{d\sigma}\psi(\sigma),
       \Omega(X)\rangle\\
       && \mbox{}-
       \int_{S^1}d\sigma\,\langle
        \ad_{\Omega\left(\varphi_{\ast}\partial_{\sigma}\right)}\psi(\sigma),
          \Omega(X)\rangle.
\end{eqnarray*}
The first term is a total derivative of a function on $S^1$; hence
vanishes.  In conclusion,
$$
\int_{S^1}d\sigma\frac{\delta\Gamma(\psi,\varphi)}{\delta\varphi}(\sigma)
 \left(X(\sigma)\right)\,=\,
 -\int_{S^1}d\sigma\langle\frac{d}{d\sigma}\psi(\sigma)+
       \ad_{\Omega\left(\varphi_{\ast}\partial_{\sigma}\right)}\psi(\sigma),
       \Omega(X)\rangle.
$$
Since we are in the smooth category and the above is true for all smooth
$X$ along $\varphi$, one must have
$$
\frac{\delta\Gamma(\psi,\varphi)}{\delta\varphi}=
    -\langle\frac{d}{d\sigma}\psi+
       \ad_{\Omega\left(\varphi_{\ast}\partial_{\sigma}\right)}\psi,
       \Omega(\cdot)\rangle.
$$
This completes the proof.

\hspace{12cm}$\Box$
%end of proof%

\medskip

{}From the previous proposition, one can now obtain the sought out
canonical transformations formally. Before doing so, we introduce
the following correspondence for necessity.

Let $\psi\in L{\frak g}$. Denote by $E(\psi)$ a path in $G$ such that
  $$
  \psi = \Omega\left(E\left(\psi\right)_{\ast}
                       \left(\partial_{\sigma}\right)\right).
  $$
{}From the theory of ordinary differential equations,
$E(\psi)$ always exists but in general doesn't close up to form a
loop. Given $\psi$, $E(\psi)$ is unique up to a left translation in $G$.

\begin{figure}
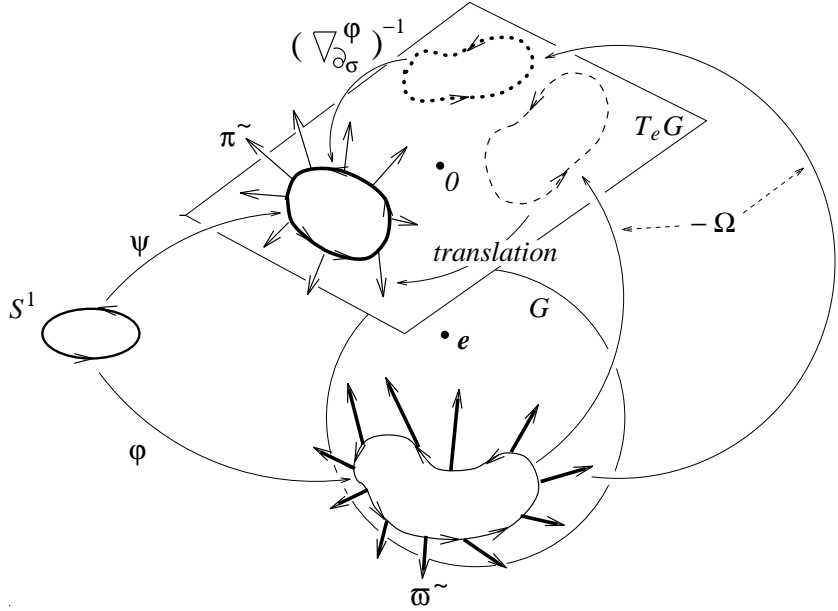

 \centerline{\figbox{dual-transform.eps}{11cm}}
\caption[dual-transform]{The formal canonical transformation between
   $LT^{\ast}G$ and $LT^{\ast}{\frak g}$.
   In the picture the metric
   dual with respect to $\langle\:,\:\rangle$
   is used to represent a 1-form along a loop. Notice that
   the tangent vector field to a loop and the momentum vector field
   along it are exchanged under the formal canonical
   transformation.}
   \protect\label{dual-transform}
\end{figure}

\begin{theorem}[Canonical transformations]
The formal canonical transformation $\Phi$ from $LT^{\ast}G$ to
$LT^{\ast}{\frak g}$ induced from the generating function $\Gamma$
is
\begin{eqnarray*}
\psi &= &-\left(\nabla_{\partial_{\sigma}}^{\varphi}\right)^{-1}
\left(\Omega\left(\varpi^{\sim}\right)\right),\\
\pi &= &-\Omega\left(\varphi_{\ast}\partial_{\sigma}\right)^{\sim},
\end{eqnarray*}
where $\nabla_{\partial_{\sigma}}^{\varphi}=\frac{d}{d\sigma}+
  \ad_{\Omega(\varphi_{\ast}\partial_{\sigma})}$ and
``$\:^{\sim}$'' represents the metric dual with respect to
$\langle\:,\:\rangle$. {\sc (Figure~\ref{dual-transform})}

Its formal inverse $\Phi^{-1}$is given by
\begin{eqnarray*}
   \varphi &= &E\left(-\pi^{\sim}\right),\\
   \varpi  &= &\left(\left(\Omega|_{\varphi}\right)^{-1}\circ
         \left(-\frac{d}{d\sigma}+\ad_{\pi^{\sim}}\right)\psi\right)^{\sim}.
\end{eqnarray*}
\end{theorem}

\medskip

\noindent
{\it Remark:} Requiring that $\psi$, $\varphi$ be loops
 puts constraint on $\varpi$ and $\pi$ respectively. Thus, like in the
 case of $S^1$, $\Phi$ and its inverse is defined only on a reduced
 phase space. One may think of this as part of certain
 ``{\it quantization conditions}''. (Or one may consider the more general
 space of ``{\it twisted loops}'' \cite{P-S}, which we will not discuss here.)

\medskip

\medskip

\noindent{\it Proof of Theorem 2.1}
The transformations are read off
straightforwardly from the previous proposition.
These transformations are
only formal due to the multi-valuedness of
operators $\left(\nabla_{\partial_{\sigma}}^{\varphi}\right)^{-1}$ and
$E$. However it turns out these multi-valuedness can be completely
understood as will be explained in the next sub-subsection.

\hspace{12cm}$\Box$
%endproof

\subsubsection{Multi-valuedness of the formal canonical transformations}
We shall show that, under our choice of generating function, the
multi-valuedness for the induced canonical transformation in either
direction is exactly what is expected.

Recall that the multi-valuedness of the map
$
E:L{\frak g}\longrightarrow \Path G
$
is parameterized by $G$ itself. Hence we only need to take care of the
multi-valuedness of the inverse,
$\left(\nabla_{\partial_{\sigma}}^{\varphi}\right)^{-1}$.

{}From the expression
$$
\nabla_{\partial_{\sigma}}^{\varphi}=\frac{d}{d\sigma}+
  \ad_{\Omega(\varphi_{\ast}\partial_{\sigma})},
$$
if we regard $\psi:S^1\longrightarrow{\frak g}$ as a section in the trivial
bundle
\begin{eqnarray*}
  S^1&\times&{\frak g}\\
   &\downarrow&\\
   &S^1&,
\end{eqnarray*}
then, given $\varphi$, the differential operator
$\nabla_{\partial_{\sigma}}^{\varphi}$ defines a
connection on this bundle.
Due to the linearity of this differential operator,
for any fixed $\varphi$, the multi-valuedness
of $\left(\nabla_{\partial_{\sigma}}^{\varphi}\right)^{-1}$ is
parameterized by the kernel
$\ker\left(\nabla_{\partial_{\sigma}}^{\varphi}\right)$.
{}From the horizontal lifting property of paths in the
base $S^1$, it must be isomorphic to a subspace of ${\frak g}$.

\begin{lemma} For any $\varphi: S^1\longrightarrow G$,
the induced connection $\nabla_{\partial_{\sigma}}^{\varphi}$ on the bundle
$S^1\times{\frak g}$ is trivial.
\end{lemma}

\noindent{\it Proof:} Let $s$ be a section in our trivial
bundle $S^1\times{\frak g}$. Observe that the following three
statements are equivalent:
\begin{eqnarray*}
   (1)&&\nabla_{\partial_{\sigma}}^{\varphi}\,s\,=\,0\:;\\
   (2)&&\frac{d}{d\sigma}s(\sigma)\,=\,
              -\ad_{\Omega(\varphi_{\ast}\partial_{\sigma})}s(\sigma)\:;\\
   (3)&&s(\sigma)\,=\,\Ad_{\varphi(\sigma)^{-1}}X_0 \hbox{ for some }
                               X_0\in{\frak g}\:.
\end{eqnarray*}

That (1) and (2) are equivalent follows from definition.

For (2) and (3), one can check that the section defined in (3)
satisfies the first-order ordinary differential equation given in (2).
Conversely, given a section $s$ that satisfies the differential equation
in (2) with $s(\sigma_0)$ specified for some $\sigma_0 \in S^1$, there
exists some $X_0\in {\frak g}$ such that
$s(\sigma_0)=\Ad_{\varphi(\sigma_0)^{-1}}X_0$ since $\Ad_g$ is an
automorphism of ${\frak g}$ for any $g\in G$.  One can then define
a new section as in (3) that coincides with $s$ at $\sigma=\sigma_0$.
Uniqueness of solutions for ordinary differential equations implies
these two sections must coincide everywhere.
Thus (2) and (3) are equivalent.

Altogether, this shows that our bundle with connection
$\nabla_{\partial_{\sigma}}^{\varphi}$ admits a combing
by globally well-defined
flat sections parametrized by the Lie algebra $\frak g$.

This concludes the proof.

\hspace{12cm}$\Box$
%endproof

\begin{corollary} The kernel
   $\ker\left(\nabla_{\partial_{\sigma}}^{\varphi}\right)$
  is isomorphic to $\frak g$.
\end{corollary}

\noindent{\it Proof:} Use the trivial connection
$\nabla_{\partial_{\sigma}}^{\varphi}$ to retrivialize the
bundle $S^1\times {\frak g}$. The kernel consists exactly
the constant sections with respect to the new trivialization
and the space of all such sections is isomorphic to $\frak g$.

\hspace{12cm}$\Box$
%endproof

\noindent{
\it Remark:} Recall that our generating function is
given by
$$
\Gamma(\psi,\varphi)=\int_{S^1}d\sigma\langle\psi(\sigma),
    \Omega(\varphi_{\ast}\partial_{\sigma})\rangle.
$$
Let $l_g$ be the left-translation by $g$.
Since $\Gamma(\psi,\varphi)=\Gamma(\psi,l_g\circ\varphi)$ for
any $g\in G$, it is expected that $\Gamma$ would determine
a canonical transformation from $LT^{\ast}G$ to
$LT^{\ast}{\frak g}$ only up to a freedom parameterized by
$G$. The corollary now shows that there is another part of freedom
parameterized by the Lie algebra. This puts the group and its
associated Lie algebra in an equal footing, which is a nice
feature as far as duality is concerned.

\bigskip

\subsection{The dual structures on the associated Lie algebra}

Continuing the previous arguments, we shall show that

\begin{theorem}
Let $G$ be a simple compact Lie group with a bi-invariant metric
$\langle\:,\:\rangle$. Let ${\frak g}$ be its associated Lie
algebra identified with $T_eG$. In order
to make the canonical transformations
worked out in the previous section be T-duality transformations,
the metric $\llangle\:,\:\rrangle$
and the $B$-field (a 2-form $B$) on ${\frak g}$ are uniquely
determined. They are given by
\begin{eqnarray*}
    \llangle X,Y\rrangle &=
        &\langle\left(\Id -\ad_v\right)^{-1}X,\left(\Id
-\ad_v\right)^{-1}Y\rangle,\\
              B(X,Y) &= &\llangle X, \ad_vY\rrangle,
\end{eqnarray*}
where $v\in{\frak g}$, $X,\,Y\in T_v{\frak g}$ and $\ad$ is the
$\ad$-representation of ${\frak g}$ on itself.
\end{theorem}

Notice that in the above expressions we implicitly identify
$T_v{\frak g}$, for any $v$ in ${\frak g}$, with ${\frak g}$ itself by
the vector space structure of ${\frak g}$.

\bigskip

\noindent{\it Proof:} We sketch first the basic ideas in the proof
and then present the details of the manipulations.

\medskip

\noindent({\it i}). {\it Basic ideas}. The inverse formal canonical
transformation $\Phi^{-1}$ from $LT^{\ast}{\frak g}$ to $LT^{\ast}G$
pulls back the string Hamiltonian function ${\cal H}$ on $LT^{\ast}G$
to some function $\widetilde{\cal H}$ on $LT^{\ast}{\frak g}$.
It turns out that this is also
a string Hamiltonian function, from which one reads off the dual metric
$\llangle\;,\;\rrangle$ and the dual $B$-field $B$ on ${\frak g}$.

\medskip

\noindent({\it ii}). {\it Details:} Recall that the Hamiltonian
${\cal H}$ on $LT^{\ast}G$ is given by
$$
{\cal H}=\int_{S^1}d\sigma {\cal H}(\varphi,\varpi;\sigma),
$$
where
$$
{\cal H}(\varphi,\varpi;\sigma)=\frac{1}{2}
\{\langle\varphi_{\ast}\partial_{\sigma},
\varphi_{\ast}\partial_{\sigma}\rangle + \langle\varpi(\sigma),
\varpi(\sigma)\rangle^{\sim}\}.
$$
To get the pulled back Hamiltonian $\widetilde{\cal H}$
on $LT^{\ast}{\frak g}$,
one simply rewrite ${\cal H}$ in terms of $(\psi,\pi)$ by using
\begin{eqnarray*}
   \varphi &= &E\left(-\pi^{\sim}\right),\\
   \varpi  &= &\left(\left(\Omega|_{\varphi}\right)^{-1}\circ
         \left(-\frac{d}{d\sigma}+\ad_{\pi^{\sim}}\right)\psi\right)^{\sim}.
\end{eqnarray*}
Now
\begin{eqnarray*}
\lefteqn{ \langle\varphi_{\ast}\partial_{\sigma},
       \varphi_{\ast}\partial_{\sigma}\rangle\;
        =\; \langle\Omega\left(\varphi_{\ast}\partial_{\sigma}\right),
       \Omega\left(\varphi_{\ast}\partial_{\sigma}\right)\rangle }\\
  & & =\; \langle -\pi^{\sim},-\pi^{\sim}\rangle
                =\langle \pi^{\sim},\pi^{\sim}\rangle
                =\langle \pi,\pi\rangle^{\sim};
\end{eqnarray*}
and
\begin{eqnarray*}
\lefteqn{ \langle\varpi,\varpi\rangle^{\sim}\;
     =\; \langle \Omega|_{\varphi}^{-1}\circ
         \left(-\frac{d}{d\sigma}+\ad_{\pi^{\sim}}\right)\psi\,,\,
        \Omega|_{\varphi}^{-1}\circ
         \left(-\frac{d}{d\sigma}+\ad_{\pi^{\sim}}\right)\psi\rangle }\\
  & & =\; \langle \left(-\frac{d}{d\sigma}+\ad_{\pi^{\sim}}\right)\psi\,,\,
            \left(-\frac{d}{d\sigma}+\ad_{\pi^{\sim}}\right)\psi\rangle\\
  & & =\; \langle \psi_{\ast}\partial_{\sigma},
                         \psi_{\ast}\partial_{\sigma}\rangle
          -2\,\langle\psi_{\ast}\partial_{\sigma},
                          \ad_{\pi^{\sim}}\psi\rangle
          + \langle \ad_{\pi^{\sim}}\psi, \ad_{\pi^{\sim}}\psi\rangle.
\end{eqnarray*}
Thus
\begin{eqnarray*}
\lefteqn{  2\widetilde{\cal H}\;
      =\; \langle \pi,\pi\rangle^{\sim}
           +\langle \psi_{\ast}\partial_{\sigma},
                  \psi_{\ast}\partial_{\sigma}\rangle
          -2\,\langle\psi_{\ast}\partial_{\sigma},
                          \ad_{\pi^{\sim}}\psi\rangle
          + \langle \ad_{\pi^{\sim}}\psi, \ad_{\pi^{\sim}}\psi\rangle }\\
  & & =\; \left(\langle \pi^{\sim},\pi^{\sim}\rangle +
        \langle \ad_{\psi}\pi^{\sim},\ad_{\psi}\pi^{\sim}\rangle\right)
     + 2\,\langle \ad_{\psi}\pi^{\sim},\psi_{\ast}\partial_{\sigma}\rangle
     + \langle \psi_{\ast}\partial_{\sigma},
                         \psi_{\ast}\partial_{\sigma}\rangle,\\
  & & \hbox{since } \ad_{\pi^{\sim}}\psi = - \ad_{\psi}\pi^{\sim}.
\end{eqnarray*}

  Next we try to put it into the string Hamiltonian form
$$
\llangle \pi-B\left(\,\cdot\,,
       \psi_{\ast}\partial_{\sigma}\right),
           \pi-B\left(\,\cdot\,,
       \psi_{\ast}\partial_{\sigma}\right)\rrangle^{\wedge}
       +\llangle\psi_{\ast}\partial_{\sigma},
                          \psi_{\ast}\partial_{\sigma}\rrangle,
$$
where $\llangle\:,\:\rrangle$ and $B$ are respectively
the sought-for metric and 2-form on ${\frak g}$
and $\llangle\:,\:\rrangle^{\wedge}$ is the induced metric on
$T^{\ast}{\frak g}$ from $\llangle\:,\:\rrangle$.

Since
$
\langle \pi^{\sim},\pi^{\sim}\rangle +
        \langle \ad_{\psi}\pi^{\sim},\ad_{\psi}\pi^{\sim}\rangle
$
contains all the quadratic terms in $\pi$, by comparison, we must have
\begin{eqnarray*}
\lefteqn{ \llangle \pi,\pi\rrangle^{\wedge}\;
        =\; \langle \pi^{\sim},\pi^{\sim}\rangle +
           \langle \ad_{\psi}\pi^{\sim},\ad_{\psi}\pi^{\sim}\rangle }\\
 & & =\; \langle (\Id +\ad_{\psi})\pi^{\sim},(\Id
+\ad_{\psi})\pi^{\sim}\rangle,
         \hbox{ since } \langle\pi^{\sim},\ad_{\psi}\pi^{\sim}\rangle = 0.
\end{eqnarray*}
Notice that the argument also implies that $\llangle\:,\:\rrangle$ is
positive definite and that, for each $\sigma$, $\Id  + \ad_{\psi(\sigma)}$
is an invertible linear transformation from
$T_{\psi(\sigma)}{\frak g} = {\frak g}$
to itself.

To get $\llangle\:,\:\rrangle$ itself, fix an orthonormal basis and its
dual basis for $({\frak g},\langle\:,\:\rangle)$. We may then regard
elements in ${\frak g}$ as a column vector and elements in ${\frak g}^{\ast}$
as a row vector.
Then, with respect to such bases, $\pi^{\sim} = \pi^{t}$, where ``$\:^{t}$''
stands for transpose; and
$$
\llangle\pi,\pi\rrangle^{\wedge}=\pi(\Id +\ad_{\psi})^{t}(\Id
+\ad_{\psi})\pi^{t}.
$$
Consequently, for $X, Y \in T_{\psi(\sigma)}{\frak g}={\frak g}$,
\begin{eqnarray*}
\lefteqn{ \llangle X,Y\rrangle\;
   =\; X^{t}\left((\Id +\ad_{\psi})^{t}(\Id +\ad_{\psi})\right)^{-1}Y }\\
 & & =\; \langle \left((\Id +\ad_{\psi})^{t}\right)^{-1}X,
                   \left((\Id +\ad_{\psi})^{t}\right)^{-1}Y\rangle\\
 & & =\; \langle (\Id -\ad_{\psi})^{-1}X,(\Id -\ad_{\psi})^{-1}Y \rangle,
\end{eqnarray*}
where we use the fact that $(\Id  + \ad_{\psi})^{t}=\Id -\ad_{\psi}$ since
$$
\langle (\Id +\ad_{\psi})X,Y\rangle = \langle X, (\Id -\ad_{\psi})Y\rangle.
$$

The mixed term
\begin{eqnarray*}
\lefteqn{ \llangle\pi,B(\:\cdot\:,
            \psi_{\ast}\partial_{\sigma})\rrangle^{\wedge} }\\
 & & =\; -\langle \ad_{\psi}\pi^{\sim},
                    \psi_{\ast}\partial_{\sigma}\rangle\;
     =\; \langle \pi^{\sim},\ad_{\psi}\psi_{\ast}\partial_{\sigma}\rangle\\
 & & =\; \langle (\Id +\ad_{\psi})\pi^{\sim},
        (\Id -\ad_{\psi})^{-1}\ad_{\psi}\psi_{\ast}\partial_{\sigma}\rangle\\
 & & =\; \langle (\Id +\ad_{\psi})\pi^{\sim},
        (\Id +\ad_{\psi})\left((\Id +\ad_{\psi})^{-1}(\Id
-\ad_{\psi})^{-1}\right)
                            \ad_{\psi}\psi_{\ast}\partial_{\sigma}\rangle\\
 & & =\; \langle (\Id +\ad_{\psi})\pi^{\sim},
   (\Id +\ad_{\psi})B(\:\cdot\:,\psi_{\ast}\partial_{\sigma})^{\sim}\rangle.
\end{eqnarray*}
Thus
$$
B\left(\:\cdot\:,\psi_{\ast}\partial_{\sigma}\right)^{\sim}
 =(\Id +\ad_{\psi})^{-1}(\Id -\ad_{\psi})^{-1}
           \ad_{\psi}\psi_{\ast}\partial_{\sigma};
$$
and
\begin{eqnarray*}
\lefteqn{ B\left(X,\psi_{\ast}\partial_{\sigma}\right) }\\
 & & =\;\langle (\Id -\ad_{\psi})^{-1}X,
     (\Id -\ad_{\psi})^{-1}\ad_{\psi}\psi_{\ast}\partial_{\sigma}\rangle\\
 & & =\;\llangle X,\ad_{\psi}\psi_{\ast}\partial_{\sigma}\rrangle.
\end{eqnarray*}
By choosing $\psi$ appropriately, we could make it pass through any
point in ${\frak g}$ along any tangent vector at that point. Thus the
above implies that
$$
B(X,Y)=\llangle X,\ad_vY\rrangle,\hbox{ for }X,Y\in T_v{\frak g}.
$$
That $B$ is a 2-form follows from the commutativity of $(\Id +\ad_v)^{-1}$,
$(\Id -\ad_v)^{-1}$, $\ad_v$ and that
$\langle X,\ad_vY\rangle = -\langle \ad_vX,Y\rangle$.

Completing the square, we then obtain
$$
2\widetilde{\cal H}=\llangle \pi-B\left(\,\cdot\,,
       \psi_{\ast}\partial_{\sigma}\right),
            \pi-B\left(\,\cdot\,,
       \psi_{\ast}\partial_{\sigma}\right)\rrangle^{\wedge}
       +\llangle\psi_{\ast}\partial_{\sigma},
                          \psi_{\ast}\partial_{\sigma}\rrangle
       + \hbox{\it Remainder},
$$
where
$$
\hbox{\it Remainder}=\langle\psi_{\ast}\partial_{\sigma},
                          \psi_{\ast}\partial_{\sigma}\rangle
  -\llangle \psi_{\ast}\partial_{\sigma},
                          \psi_{\ast}\partial_{\sigma}\rrangle
  -\llangle B\left(\,\cdot\,,
       \psi_{\ast}\partial_{\sigma}\right),
            B\left(\,\cdot\,,
       \psi_{\ast}\partial_{\sigma}\right)\rrangle^{\wedge}.
$$

We shall now show that $\hbox{\it Remainder }=0$.
\begin{eqnarray*}
\lefteqn{ \mbox{\it Remainder}}\\
 & & =\;\langle\psi_{\ast}\partial_{\sigma},
                          \psi_{\ast}\partial_{\sigma}\rangle
         -\langle (\Id -\ad_{\psi})^{-1}\psi_{\ast}\partial_{\sigma},
           (\Id -\ad_{\psi})^{-1}\psi_{\ast}\partial_{\sigma}\rangle \\
 & &\mbox{}
     -\langle (\Id -\ad_{\psi})^{-1}\ad_{\psi}\psi_{\ast}\partial_{\sigma},
     (\Id -\ad_{\psi})^{-1}\ad_{\psi}\psi_{\ast}\partial_{\sigma}\rangle\\
 & & =\; \langle \psi_{\ast}\partial_{\sigma},
             [\Id -(\Id +\ad_{\psi})^{-1}(\Id -\ad_{\psi})^{-1}
            +(\Id +\ad_{\psi})^{-1}(\Id -\ad_{\psi})^{-1}\ad_{\psi}^2]
                 \psi_{\ast}\partial_{\sigma}\rangle\\
 & & =\; \langle \psi_{\ast}\partial_{\sigma},
           (\Id -\ad_{\psi}^2)^{-1}[(\Id -\ad_{\psi}^2)-\Id +\ad_{\psi}^2]
                \psi_{\ast}\partial_{\sigma}\rangle\\
 & & =\; 0 \mbox{ as claimed}.
\end{eqnarray*}

{}From the argument it is clear that
the $\llangle\:,\:\rrangle$ and $B$ are uniquely
determined by the formal canonical transformation.

This completes the proof.

\hspace{12cm}$\Box$
%endproof

\subsection{A second glance at $\Phi$}

As already pointed out in a remark following Theorem 2.1, the formal
canonical transformation $\Phi$ that we constructed from the
generating function $\Gamma$ is not a map from the whole $LT^{\ast}G$
to the whole $LT^{\ast}{\frak g}$. Instead, it singles out a reduced
phase-space, $\Dom \Phi$, the domain of $\Phi$ in $LT^{\ast}G$ and
a reduced phase-space, $\Im \Phi$, the image of $\Phi$ in
$LT^{\ast}{\frak g}$. Both are of codimension $\dim  G$
in the original phase-spaces they reside. One last question concerning
the target-space duality between $(G,\langle\:,\:\rangle)$ and
$({\frak g},\llangle\:,\:\rrangle,B)$ at the classical level
under the Hamiltonian formalism based on $\Gamma$ is then:
$$
\mbox{\it Q. Are both $\Dom\Phi$ and $\Im \Phi$ invariant
under the string Hamiltonian flows?}
$$
Naively, one would expect that
if $\Dom \Phi$ is invariant under the string Hamiltonian
flow in $LT^{\ast}G$, then so is $\Im \Phi$ in $LT^{\ast}{\frak g}$
due to the way the string Hamiltonian $\widetilde{\cal H}$ on
$LT^{\ast}{\frak g}$ is constructed. Also notice that, a priori,
the domain and image of a canonical transformation that a generating
function generates do not necessarily have to do with Hamiltonian flows.
However, as a thumb rule that whatever is natural tends to work, the
answer to the above question is affirmative. This certainly gives
another backup of the Hamiltonian setting presented here and in the
literature.

\begin{theorem}[Invariance under string Hamiltonian flow]
 Both $\Dom\Phi$ and $\Im\Phi$ are invariant under the related
 string Hamiltonian flow. Thus they do form sub- string Hamiltonian
 systems.
\end{theorem}

\noindent{\it Proof}.
Recall from Theorem 2.1 the map $E$ from $L{\frak g}$ to $\Path G$
and the connection $\nabla_{\partial_{\sigma}}^{\varphi}$.
These together with the proof of Lemma 2.1 leads to
\begin{eqnarray*}
\lefteqn{\Dom\Phi\;=\;\left\{(\varphi,\varpi)\left|\,
       \varphi\,\in\,LG,\;\mbox{ and }\,\Omega(\varpi^{\sim})\,
        \in\,\Im \nabla_{\partial_{\sigma}}^{\varphi}.\right.\right\}  }\\
 & & =\;\left\{(\varphi,\varpi)\left|
       \varphi\,\in\,LG,\;\mbox{ and }\,\int_{S^1}d\sigma\,
        \Ad_{\varphi(\sigma)}\,\Omega(\varpi^{\sim})(\sigma)\:=\:0.
                                                        \right.\right\};
\end{eqnarray*}
and
$$
\Im\Phi\;=\;\left\{(\psi,\pi)\left|\,\psi\,\in\,L{\frak g},\;\mbox{ and }\,
       E(-\pi^{\sim})\,\mbox{ is a loop in $G$.}\right.\right\}.
$$
We shall first show that $\Dom \Phi$ is invariant under the flow
generated by the string Hamiltonian vector field $X_{\cal H}$.
We begin with a
condition that characterizes $T_{\ast}(\Dom \Phi)$ and then show that
$X_{\cal H}$, when restricted to $\Dom \Phi$, satisfies this
condition and hence has to be tangent to $\Dom \Phi$. The flow
generated therefore leaves $\Dom \Phi$ invariant. The invariance
of $\Im \Phi$ under the flow generated by $X_{\widetilde{\cal H}}$
is also demonstrated by a similar approach.

Let us introduce the following trivializations of bundles in the
discussion:
$$
T_{\ast}G\,\stackrel{\Omega}{\simeq}\,G\times {\frak g},\hspace{0.5cm}
T^{\ast}G\,\stackrel{\Omega}{\simeq}\,G\times{\frak g}^{\ast},
\hspace{0.5cm} LT^{\ast}G\,\stackrel{\Omega}{\simeq}\,
                                     LG\times L{\frak g}^{\ast},
$$
$$
T_{\ast}(LT^{\ast}G)\,\stackrel{\Omega}{\simeq}
      T_{\ast}LG\times T_{\ast}L{\frak g}^{\ast},\hspace{0.5cm}
T^{\ast}(LT^{\ast}G)\,\stackrel{\Omega}{\simeq}
      T^{\ast}LG\times T^{\ast}L{\frak g}^{\ast}.
$$
We denote all these bundle isomorphisms by $\Omega$ since they
all arise from the first isomorphism which defines the Maurer-Cartan
form $\Omega$.

Let $(\varphi_t,\varpi_t)$ be a path in $LT^{\ast}G$ that lies in
$\Dom \Phi$ with
$$
\Omega\left(\left.\frac{d}{dt}\right|_{t=0}(\varphi_t,\varpi_t)\right)\;
  =\;(Y,Z).
$$
Then
$$
\int_{S^1}d\sigma\,\Ad_{\varphi_t(\sigma)}\,
   \Omega\left(\varpi_t^{\sim}(\sigma)\right)\;=\;0\;\;
    \mbox{ for all $t$.}
$$
Thus
\begin{eqnarray*}
\lefteqn{0\;=\;\left.\frac{d}{dt}\right|_{t=0}\,\int_{S^1}d\sigma\,
 \Ad_{\varphi_t(\sigma)}\,\Omega\left(\varpi_t^{\sim}(\sigma)\right) }\\
 & & =\;\int_{S^1}d\sigma\,\Ad_{\varphi_0(\sigma)}\,
       \left.\frac{\partial}{\partial t}\right|_{t=0}
        \left( \Ad_{\varphi_0(\sigma)^{-1}\varphi_t(\sigma)}\,
         \Omega\left(\varpi_t^{\sim}(\sigma)\right)\right)\\
 & & =\;\int_{S^1}d\sigma\,\Ad_{\varphi_0(\sigma)}\,
         \left\{ \ad_{Y(\sigma)}\,
          \Omega\left(\varpi_0^{\sim}(\sigma)\right)\,
                                +\,Z(\sigma)^{\sim}\right\}.
\end{eqnarray*}
And we lead to a criterion for a tangent vector
$(Y,Z)$ at $(\varphi,\varpi)$ in $\Dom \Phi$
to be in $T_{\ast}(\Dom \Phi)$:
$$
\int_{S^1}d\sigma\,\Ad_{\varphi(\sigma)}\,
         \left\{ \ad_{Y(\sigma)}\,
          \Omega\left(\varpi^{\sim}(\sigma)\right)\,
                         +\,Z(\sigma)^{\sim}\right\}\;=\;0.
$$
(One should think of $Y$ as an arbitrary smooth vector field along
$\varphi$ in $G$; and then $Z$ is subject to the above constraint.)

Next recall that the string Hamiltonian $\cal H$ on $LT^{\ast}G$ has
density
$$
{\cal H}(\varphi,\varpi;\sigma)\;
   =\;\frac{1}{2}\langle\varpi(\sigma),\varpi(\sigma)\rangle^{\sim}\,
     +\,\frac{1}{2}\langle \varphi_{\ast}\partial_{\sigma},
            \varphi_{\ast}\partial_{\sigma}\rangle,
$$
from which one has
\begin{eqnarray*}
\lefteqn{ d{\cal H}(\varphi,\varpi;\sigma)\;
   =\;\langle\varpi(\sigma),\:\cdot\:\rangle^{\sim}\,
     -\,\langle\nabla_{\varphi_{\ast}\partial_{\sigma}}
              \varphi_{\ast}\partial_{\sigma},\:\cdot\:\rangle }\\
 & & =\;\varpi^{\sim}(\sigma)\,
       -\,\left(\nabla_{\varphi_{\ast}\partial_{\sigma}}
              \varphi_{\ast}\partial_{\sigma}\right)^{\sim}.
\end{eqnarray*}
Consequently,
$$
\Omega\left(d{\cal H}(\varphi,\varpi)\right)(\sigma)\;
  =\;\left(-\Omega\left(\left(\nabla_{\varphi_{\ast}\partial_{\sigma}}
              \varphi_{\ast}\partial_{\sigma}\right)^{\sim}\right)\,,
    \,\Omega\left(\varpi^{\sim}\right)(\sigma)\right);
$$
and
$$
\Omega\left(\left.X_{\cal H}\right|_{(\varphi,\varpi)}\right)(\sigma)\;
  =\;\left( \Omega(\varpi^{\sim}(\sigma))\,,\,
    \Omega\left((\nabla_{\varphi_{\ast}\partial_{\sigma}}
                   \varphi_{\ast}\partial_{\sigma})^{\sim}\right)\right).
$$

Now we only need to check that
$\Omega\left(\left.X_{\cal H}\right|_{(\varphi,\varpi)}\right)$
satisfies the above criterion for $(\varphi,\varpi)$ in $\Dom \Phi$.

\begin{eqnarray*}
\lefteqn{\int_{S^1}d\sigma\,\Ad_{\varphi(\sigma)}\,\left\{
  \ad_{\Omega(\varpi^{\sim}(\sigma))}\Omega(\varpi^{\sim}(\sigma))\,
   +\,\Omega(\nabla_{\varphi_{\ast}\partial_{\sigma}}
                            \varphi_{\ast}\partial_{\sigma})\right\} }\\
 & & =\;\int_{S^1}d\sigma\,\Ad_{\varphi(\sigma)}\,
        \frac{d}{d\sigma}\Omega(\varphi_{\ast}\partial_{\sigma})\\
 & & \hspace{2cm}\mbox{ since $\ad_YY\,=\,0$ and
       $\Omega(\nabla_{\varphi_{\ast}\partial_{\sigma}}
                            \varphi_{\ast}\partial_{\sigma})\:
        =\:\frac{d}{d\sigma}\Omega(\varphi_{\ast}\partial_{\sigma})$  };\\
 & & =\;-\,\int_{S^1}d\sigma\,
       \left(\frac{d}{d\sigma}\Ad_{\varphi(\sigma)}\right)\,
          \Omega(\varphi_{\ast}\partial_{\sigma})\\
 & & =\;-\,\int_{S^1}d\sigma\,\Ad_{\varphi(\sigma)}\,
            \ad_{\Omega(\varphi_{\ast}\partial_{\sigma})}
                   \Omega(\varphi_{\ast}\partial_{\sigma})\; =\;0.
\end{eqnarray*}
This shows the invariance of $\Dom \Phi$ under the flow generated by
$X_{\cal H}$.

Likewise for the image $\Im \Phi$  we shall introduce the trivialization
of the respective bundles induced by the trivialization
$$
T_{\ast}{\frak g}\,=\,{\frak g}\times {\frak g},
$$
arising from the vector space structure of ${\frak g}$.
A similar argument as in the first part, using the identity
$$
\left.\frac{\partial}{\partial t}\right|_{t=0}
       \Omega\left(\varphi_{t\ast}\partial_{\sigma}\right)\;
   =\; \frac{d}{d\sigma}\Omega(T_0)
     + \ad_{\Omega\left(\varphi_{\ast}\partial_{\sigma}\right)}\Omega(T_0)
$$
in the proof of Proposition 2.1 with $T_0(\sigma)$ being
$\left.\frac{\partial}{\partial t}\right|_{t=0}\varphi_{t}(\sigma)$,
gives the following criterion for $T_{\ast}\Im \Phi$:
$$
(Y,Z)\,\in\,T_{(\psi,\pi)}\Im \Phi\hspace{.5cm}\mbox{\it iff}
\hspace{.5cm}Z^{\sim}\,\in\,
   \Im \left(-\frac{d}{d\sigma} + \ad_{\pi^{\sim}}\right).
$$
It remains to show that $X_{\widetilde{\cal H}}$, when restricted to
$\Im \Phi$ satisfies this criterion. Let
$$
X_{\widetilde{\cal H}}\,=\,(Y_{\widetilde{\cal H}},Z_{\widetilde{\cal H}}).
$$
Then in terms of a parallel orthonormal frame in $\frak g$ with respect
to $\langle\:,\:\rangle$, one has
$$
X_{\widetilde{\cal H}}\:=\:(Y_{\widetilde{\cal H}},
  Z_{\widetilde{\cal H}})\:
  =\:\left(\frac{\delta\widetilde{\cal H}}{\delta\pi},
          -\frac{\delta\widetilde{\cal H}}{\delta\psi}\right).
$$
Since only the $Z$-component matters, we shall work the latter
functional derivative out.

Recall from the proof of Theorem 2.2 that
\begin{eqnarray*}
\widetilde{\cal H}(\psi,\pi)
 &=&\frac{1}{2}\,\int_{S^1} d\sigma\,\bigg\{
     \langle\pi^{\sim},\pi^{\sim}\rangle
    + \langle \ad_{\psi}\pi^{\sim},\ad_{\psi}\pi^{\sim}\rangle\\
    &&\mbox{}+ 2\,\langle \ad_{\psi}\pi^{\sim},
               \psi_{\ast}(\partial_{\sigma})\rangle
    + \langle\psi_{\ast}\partial_{\sigma},
                       \psi_{\ast}\partial_{\sigma}\rangle \bigg\}.
\end{eqnarray*}
Let $\psi_t$ be a path in $L{\frak g}$ with $\psi_0\,=\,\psi$ and
$\left.\frac{d}{dt}\right|_{t=0}\,\psi_t\:=\:T_0$. Then
\begin{eqnarray*}
 \lefteqn{\left.\frac{d}{dt}\right|_{t=0}\,
                 \widetilde{\cal H}(\psi_t,\pi)  }\qquad \\
 &= & \int_{S^1} d\sigma\,\bigg\{
           \langle \ad_{T_0}\pi^{\sim},\ad_{\psi}\pi^{\sim}\rangle
         + \langle \ad_{T_0}\pi^{\sim},
               \psi_{\ast}\partial_{\sigma}\rangle \\
    &&\mbox{}     + \langle \ad_{\psi}\pi^{\sim},
               \nabla_{T_0}\psi_{t\ast}\partial_{\sigma}\rangle
         + \langle\nabla_{T_0}\psi_{t\ast}\partial_{\sigma},
                  \psi_{t\ast}\partial_{\sigma}\rangle \bigg\} \\
 & & \mbox{(where $\nabla$ is the connection associated to
            the flat metric $\langle\:,\:\rangle$ on $\frak g$)}\\
 &= & \int_{S^1} d\sigma\,\left\langle T_0\:,\:
         \ad_{\pi^{\sim}}\ad_{\psi}\pi^{\sim}\,
       +\, \ad_{\pi^{\sim}}\psi_{\ast}\partial_{\sigma}\,
       -\, \frac{d}{d\sigma}\left(\ad_{\psi}\pi^{\sim}\right)\,
       -\, \frac{d^2}{dt^2}\psi \right\rangle.
\end{eqnarray*}
Thus, with respect to the same orthonormal parallel frame,
\begin{eqnarray*}
 \left(Z_{\widetilde{\cal H}}\right)^{\sim}\;
    &=& \left(\ad_{\pi^{\sim}}\right)^2\psi\,
        -\,\ad_{\pi^{\sim}}\psi_{\ast}\partial_{\sigma}\,
        -\,\frac{d}{d\sigma}\left(\ad_{\pi^{\sim}}\psi\right)
        +\,\frac{d^2}{d\sigma^2}\psi  \\
 &+ & \left(-\frac{d}{d\sigma}+\ad_{\pi^{\sim}}\right)
   \left( \ad_{\pi^{\sim}}\psi - \psi_{\ast}\partial_{\sigma} \right),
\end{eqnarray*}
which satisfies the criterion for tangency to $\Im \Phi$.

This completes the proof.

\hspace{12cm}$\Box$
\medskip

\par
\begin{figure}
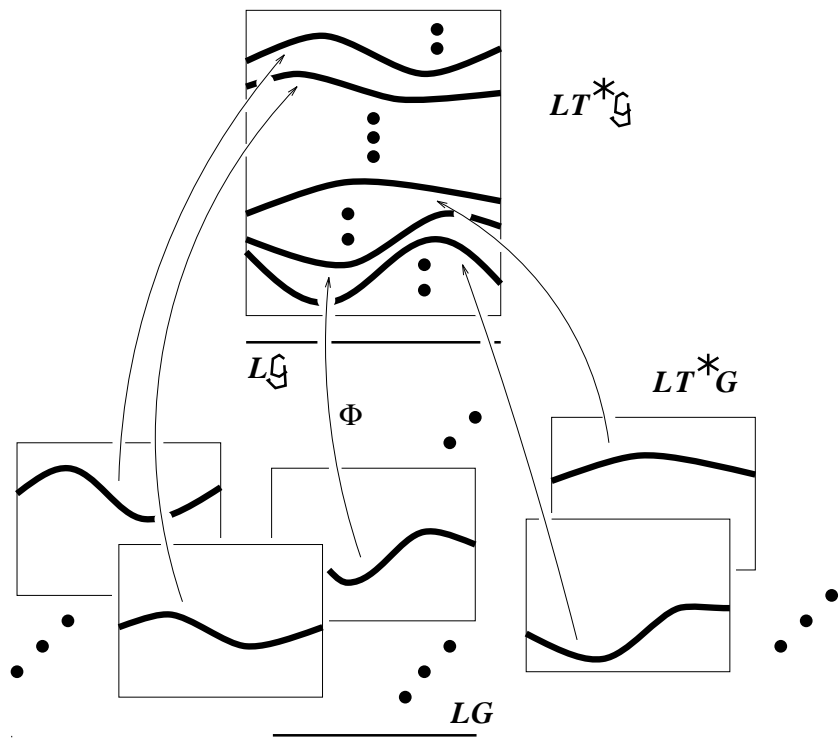

 \centerline{\figbox{phi.eps}{11cm}}
 \caption[Phi-fig]{The rectangles on the bottom represent the components
 of $LT^*G$ and the thick curves are
 the components of $\Dom\Phi$. On the top rectangle, the thick curves
 represent the components of $\Im\Phi$.}
 \protect\label{Phi-fig}
\end{figure}

\noindent{\it Remark:}
{}From the conditions that characterizes $\Dom\Phi$ and $\Im \Phi$, one
can see that:
\begin{enumerate}
\item  $\Dom\Phi$ is a codimension $\dim  G$ vector
subbundle in $LT^{\ast}G$ over $LG$;
it has one component over each component of $LG$
and these components are labeled exactly by $\pi_1(G)$ because
$\pi_0(LG)=\pi_1(G)$.
\item $\Im \Phi$
is also a codimension $\dim  G$ subspace in $LT^{\ast}{\frak g}$; it is
a bundle over $L{\frak g}$ with components of the fiber parameterized
again by $\pi_1(G)$.
\item $\Phi$ then takes a component of $\Dom \Phi$ onto
a component of $\Im \Phi$ ({\sc Figure~\ref{Phi-fig}}).
\end{enumerate}
This indicates that analogously to the $S^1$ case, $\Phi$ cannot be extended
to a bijection between the two unreduced phase spaces after factoring out
the redundancy.

\medskip

\par\noindent
{\it Remark:}
In comparison with the $S^1$ (i.e. $U(1)$) case, some features are
similar and other features are missing.
 Actually the $S^1_R$ - $S^1_{\frac{1}{R}}$
T-dual pair can be obtained from the general setting with
the additional introduction of an
appropriate compactification of the associated Lie algebra $\frak g$.
Naively, when this is done, say by a lattice in $\frak g$,
one may then extend the scope from the loop space to the space of paths
with the difference of end-points lying in the lattice.
In the case of  ${\frak u}(1)$ there is no difficulty,
however, for general simple Lie group - Lie algebra pairs,
it is not clear how such a compactification can be introduced that
lives compatibly with all other properties.

\medskip

So far, we have rarely touched upon the symmetries in the theory.
A reason for this is that many of the concepts and discussions used
here are quite general. Actually, one may see that all the arguments
seem to be applicable even to cases without symmetries as long as
one is able to tell what is the generating function.   We
hope that using arguments which avoid relying heavily on  symmetries may
shed some light on the more challenging situations.
Nevertheless, it is worthwhile to see if symmetry provides any
conceptual reason why things should work in the present case.

\subsection{Symmetries in the theory}

Conceptually and naively, the following may be related:
\begin{itemize}
 \item symmetries of the generating function $\Gamma$;
 \item the redundancy (i.e. the non-injectiveness and multi-valuedness)
       of $\Phi$;
 \item symmetries of the Hamiltonian systems for the $G$-part and
       the $\frak g$-part respectively.
\end{itemize}
We shall try to clarify their relationships with each other.

\subsubsection{Symmetries of $\Gamma$ and the redundancy of $\Phi$}

A more symmetric way to think of $\Phi$ is to regard it as a
{\it symplectic relation}. The section $d{\Gamma}$ in
$T^{\ast}(L{\frak g}\times LG)$ over $L{\frak g}\times LG$ gives rise
to an embedded Lagrangian submanifold in the product space
$LT^{\ast}{\frak g}\times LT^{\ast}G$ with the symplectic
structure
$\tilde{\mbox{\boldmath $\omega$}}\ominus\mbox{\boldmath $\omega$}$;
this then leads to a relation from $LT^{\ast}G$ to $LT^{\ast}{\frak g}$,
which is exactly $\Phi$. Let $\Sym _{\Gamma}$ be a group acting on
$L{\frak g}\times LG$ that leaves $\Gamma$ invariant. Then its induced
action on $LT^{\ast}{\frak g}\times LT^{\ast}G$ is symplectic and
leaves $d\Gamma$ invariant. The intersection of $\Sym _{\Gamma}$-orbits in
$LT^{\ast}{\frak g}\times LT^{\ast}G$ with the vertical leaves of
the product space then contributes to the non-injectiveness of $\Phi$,
while that with the horizontal leaves contributes to the
multi-valuedness of $\Phi$. This gives a general picture how the
symmetry of $\Gamma$ and the redundancy of $\Phi$ are related.

In the present case, there are at least two groups of symmetries for
$\Gamma$:
$$
(\Sym _{\Gamma})_1\;=\;L{\frak g}
  \hspace{.5cm}\mbox{with the pointwise addition operation from $\frak g$},
$$
whose action on $L{\frak g}\times LG$ is defined by
$$
\begin{array}{ccc}
 (\Sym _{\Gamma})_1\times(L{\frak g}\times LG)
                         & \longrightarrow   & L{\frak g}\times LG\\
  (\eta,\psi,\varphi)  & \longmapsto
  &(\psi + \ad_{\Omega(\varphi_{\ast}\partial_{\sigma})}\eta\,,\,\varphi)\;;
\end{array}
$$
and
$$
(\Sym _{\Gamma})_2\;=\; G_L\times G_R
  \hspace{.5cm}\mbox{with the componentwise
                               multiplication from that of $G$},
$$
whose action on $L{\frak g}\times LG$ is defined by
$$
\begin{array}{ccc}
 (\Sym _{\Gamma})_2\times(L{\frak g}\times LG)
                         & \longrightarrow   & L{\frak g}\times LG\\
 \left((g_1,g_2),\psi,\varphi\right)  & \longmapsto
  & (\Ad_{g_2^{-1}}\psi, l_{g_1}r_{g_2}\varphi)\;.
\end{array}
$$
Direct computation shows that the intersection of an
$L{\frak g}$-orbit in $LT^{\ast}{\frak g}\times LT^{\ast}G$
with either a vertical or a horizontal leaf is in general just a point;
hence this huge symmetry of $\Gamma$ actually won't contribute to the
redundancy of $\Phi$ in a major way. However the $G_L\times G_R$-orbit
of a point in $LT^{\ast}{\frak g}\times LT^{\ast}G$ intersects the
vertical leaf through that point by the $G_L\times{e_R}$-suborbit,
which is homeomorphic to $G$. Thus the $G_L\times G_R$-symmetry of
$\Gamma$ accounts for the non-injectiveness of $\Phi$ completely.
On the other hand, the same orbit intersects the horizontal leaf through
that point at only one point and, hence, this action doesn't
contribute to the multi-valuedness of $\Phi$.

\subsubsection{Symmetries of the string Hamiltonian systems}

{\bf (a) The $G$-part:} Since the metric
$\langle\:,\:\rangle$ on $G$ is bi-invariant, the Hamiltonian system
$(LT^{\ast}G,{\cal H})$ admits a $G_L\times G_R$ action induced
by the left- and right- multiplication in $G$.
Moreover, since this action preserves the canonical symplectic potential
{\boldmath $\theta$} on $LT^{\ast}G$, there is a {\it moment map}
\cite{A-M}
$$
\mu\,=\,(\mu_L,\mu_R)\::\:LT^{\ast}G\:\longrightarrow\:
     {\frak g}_L^* \oplus{\frak g}_R^*
$$
defined by
\begin{eqnarray*}
\lefteqn{\mu(\varphi,\varpi)(v_1,v_2)\;
  =\; (\mu_L(\varphi,\varpi)(v_1)\,,\,\mu_R(\varphi,\varpi)(v_2)) }\\
 & &  =\; \left( \int_{S^1}d\sigma\,
        \varpi(\sigma)(\xi^L_{v_1}|_{\varphi(\sigma)})\,,\,
     \int_{S^1}d\sigma\,
       \varpi(\sigma)(\xi^R_{v_2}|_{\varphi(\sigma)})\right)\,,
\end{eqnarray*}
where $\xi^L_{v_1}$ (resp. $\xi^R_{v_2}$) is the left (resp. right)
invariant vector field on $G$ generated by $v_1$ (resp. $v_2$).

\begin{proposition} $\Dom \Phi\;=\;\mu_R^{-1}(0)$.
\end{proposition}

\noindent
{\it Proof}. This follows from the computation:
\begin{eqnarray*}
\lefteqn{\mu_R(\varphi,\varpi)(v)\;
  =\; \int_{S^1}d\sigma\,
       \varpi(\sigma)(\xi^R_{v}|_{\varphi(\sigma)})  }\\
 & & =\; \int_{S^1}d\sigma\, \langle\varpi(\sigma)^{\sim}\,,\,
           \xi^R_{v}|_{\varphi(\sigma)}\rangle\;
     =\; \int_{S^1}d\sigma\, \langle\Omega(\varpi(\sigma)^{\sim})\,,\,
                  \Ad_{\varphi(\sigma)^{-1}}v\rangle  \\
 & & =\; \int_{S^1}d\sigma\,\langle
        \Ad_{\varphi(\sigma)}\Omega(\varpi(\sigma)^{\sim})\,,\,v\rangle\;
     =\;\langle \int_{S^1}d\sigma\,
          \Ad_{\varphi(\sigma)}\Omega(\varpi(\sigma)^{\sim})\,,\,v\rangle\,.
\end{eqnarray*}
This vanishes for all $v$ iff
$$
\int_{S^1}d\sigma\,
  \Ad_{\varphi(\sigma)}\Omega(\varpi(\sigma)^{\sim})\,=\,0\,,
$$
which is exactly the condition that characterizes $\Dom \Phi$.

\hspace{12cm}$\Box$

This proposition suggests that one may apply the {\it Marsden-Weinstein
reduction} to $(LT^{\ast}G,{\cal H})$ and consider the quotient space
$\Dom \Phi/G_R$ as the true classical physical phase space. It also
provides a ``{\it true}'' reason for the invariance of $\Dom \Phi$ under
the flow generated by $X_{\cal H}$.

\medskip

\noindent
{\bf (b) The $\frak g$-part:}

The identity
$$
\Gamma(\psi, l_{g_1}r_{g_2}\varphi)\;=\;\Gamma(\Ad_{g_2}\psi,\varphi)
$$
suggests that $\Phi$ transforms the $G_L\times G_R$ action on
$LT^{\ast}G$ into the $\Ad G_R$ action on $LT^{\ast}{\frak g}$.
Indeed as will be shown in Sec.~3.3, the T-dual Hamiltonian system
$(LT^{\ast}{\frak g},\widetilde{\cal H})$ does admit the $\Ad G$ action.
Analogous to the $G$-part, this also leads to a moment map
$$
\tilde{\mu}\;:\:LT^{\ast}{\frak g}\:\longrightarrow\:{\frak g}^*
$$
defined by
$$
\tilde{\mu}(\psi,\pi)(v)\;=\;\int_{S^1}d\sigma\,\pi(\sigma)
   (\eta_v|_{\psi(\sigma)}),
$$
where $\eta_v$ is the vector field on $\frak g$ associated to $v$
via the $\Ad$-action. Explicitly, $\eta_v|_{\psi(\sigma)}$ is just
$\ad_v\psi(\sigma)$. Direct computation then gives
$$
\tilde{\mu}(\psi,\pi)(v)\; =\; \langle \int_{S^1}d\sigma\,
   \ad_{\psi(\sigma)}\pi^{\sim}(\sigma)\,,\,v\rangle.
$$
Unfortunately, we are not able to see if $\Im \Phi$ is of the form
$\tilde{\mu}^{-1}(A)$ for some subset $A$ in $\frak g$ by using the
above formula. The condition that characterizes $\Im \Phi$ is more
related to the following {\it holonomy map}
$$
\begin{array}{ccccc}
\Hol & : & LT^{\ast}{\frak g} & \longrightarrow & G  \\
    & & (\psi,\pi) & \longmapsto &
          E(-\pi^{\sim})_0^{-1}E(-\pi^{\sim})_{2\pi}\,.
\end{array}
$$
Notice that this is well-defined regardless of the initial point
$E(-\pi^{\sim})_0$ chosen. It is not clear to us how to translate
the fact that $\Im\Phi=\Hol^{-1}(e)$ into the language of
$\tilde{\mu}$. There might be other symmetries that would
give the correct moment map for such a translation.
And we shall conclude our discussion of symmetries with this open end.

The last issue that we shall touch upon in this article is about the
T-dual structures on $\frak g$. There are surely many more
properties worth studying, in particular, the curvature properties,
the asymptotic behavior of geodesics of the T-dual Riemannian
manifold and the existence of symplectic leaves of $B$.
However we shall be contented here only to give a light feel
of the T-dual geometry on $\frak g$.

\section{The Geometry of $(\mbox{\bigfrak g},\llangle\:,\:\rrangle\,B)$}

\subsection{Preliminaries to the study the dual geometry}

The dual structures on $\frak g$ worked out in the previous section
links closely to the $\ad$-representation of $\frak g$ on itself.
Thus in this sub-section, we shall digress to prepare ourselves
necessary facts about real simple Lie algebras and
their $\ad$-representation for studying the dual geometry.
These facts either are contained in \cite{Sa} or can be derived from
material therein.

\subsubsection{The characteristic polynomials and the characteristic variety}

For any $v\,\in\,{\frak g}$, the characteristic
polynomial is defined to be
$$
\det\left(\,\ad_v\,-\,t\Id\right) = (-1)^{n}\left(t^n-D_1(v)t^{n-1}
        +D_2(v)t^{n-2}-\cdots+(-1)^{n-r}D_{n-r}(v)t^r\right),
$$
where $n\,=\,\dim {\frak g}$ and $r\,=\,\rank {\frak g}$.
Notice that the coefficients $D_i$ are homogeneous polynomials in $v$ of
degree $i$.

Since $\frak g$ is simple, one has that $n-r$ is even and that
the characteristic polynomials can be written in the form
$$
\det(\ad_v-t\Id)=(-1)^n t^r
\prod_{i=1}^{\frac{n-r}{2}}\left(t^2+a_i(v)^2\right),
$$
where $a_i$ are some functions in $v$.
This implies that
$$
D_i\equiv 0\;,\hbox{ for $i$ odd}.
$$

The characteristic variety $V_0$ is defined to be the zero set of the
homogeneous polynomial $D_{n-r}$. It is naturally stratified by the
following ``tower''
$$
V_0\supset V_1\supset \cdots \supset V_k \supset \cdots
\supset V_{\frac{n-r}{2}-1}(= \{0\}),
$$
where
$$
V_k = \{v\mid D_{n-r}(v)\,=\,D_{n-r-2}(v)\,=\,\cdots\,=\,D_{n-r-2k}(v)
\,=\,0\}.
$$

\subsubsection{Cartan subalgebras}

\begin{itemize}
  \item Let $v\in {\frak g}$. Then there exists exactly one Cartan
        subalgebra that contains $v$ if and only  if $v\in {\frak g}-V_0$.
  \item If $v\in V_k$ for $k > 0$, then there exists at least a
        $2(k+1)$-dimensional family of Cartan subalgebras and each
        contains $v$.
  \item Fix a Cartan subalgebra ${\frak h}$ and an
        $\Ad$-invariant inner product $\langle\;,\;\rangle$ in $\frak g$.
        Let $\Delta$ be the set of roots of $\frak g$ with respect to
        $\frak h$ with a fixed order. Then $\frak g$ decomposes orthogonally
        into
        $$
        {\frak g}={\frak h}\oplus\left(\oplus_{\alpha\in\Delta^{+}}
        \Pi_{\alpha}\right),
        $$
        where $\Delta^{+}$ is the set of all positive roots and each
        $\Pi_{\alpha}$ is a 2-dimensional subspace invariant under
        $\ad{\frak h}$.
  \item For any $v\in {\frak g}$, the kernel of the endomorphism
        $\ad_v:{\frak g}\to{\frak g}$ contains all
        the Cartan subalgebras that contain $v$.
\end{itemize}

\subsubsection{The $\Ad$-action on $\frak g$ and the Weyl group}

\begin{itemize}
 \item The $\Ad$-action of $G$ on $\frak g$ induces a $G$-action on
       the space of all Cartan subalgebras (with the subset topology
       from an appropriate Grassmann manifold). This induced action
       is transitive.
 \item Since $\langle\;,\;\rangle$ is $\Ad$-invariant and $\frak g$ is
       compact simple, one has a group homomorphism
       $$
       \Ad\,:\,G\,\longrightarrow\,SO(n)\subset
        \Isom({\frak g},\langle\;,\;\rangle).
       $$
 \item Let $T$ be the maximal torus in $G$ associated to $\frak h$, i.e.
       ${\frak h}=T_eT$. The restricted action $\Ad T$ on $\frak g$ leaves
       $\frak h$ fixed and are rotations on each $\Pi_{\alpha}$.
 \item Let $h_{\alpha}$ be root vectors associated to roots
       $\alpha\in\Delta$. Recall that the Weyl group action $\cal W$
       on $\frak h$
       associated to $\Delta$ is generated by the reflections with
       respect to $h_{\alpha}^{\bot}$, the orthogonal complement of
       $h_{\alpha}$ in $\frak h$ with respect to $\langle\;,\;\rangle$.
       Then every element in $\cal W$ comes from an $\Ad_g$, for some
       $g\in G$, that leaves $\frak h$ invariant. Conversely, if
       some $\Ad_g$, $g\in G$, leaves $\frak h$ invariant, then the
       restriction $\Ad_g\mid_{\frak h}$ is in $\cal W$.
\end{itemize}

\subsubsection{The Weyl chamber}

For our purposes, the {\it Weyl chamber} associated to $({\frak h},\Delta)$
with a fixed order  shall mean any of the following:
\begin{itemize}
  \item The closed Weyl chamber is the quotient space ${\frak h}/{\cal W}$.
        It is a convex cone with boundary. The interior is called the
        open Weyl chamber.
  \item Let $F$ be a fixed fundamental system. Then the closed Weyl
        chamber is the cone
        $$
        C=C_F=\{v\in{\frak h}\mid \langle h_{\alpha},v\rangle\ge 0\}.
        $$
        Its interior is the open Weyl chamber.
  \item Let $\Sigma_{\alpha}=\{ v\in{\frak h}\mid
            \langle h_{\alpha},v\rangle=0\}$, for $\alpha\in\Delta$.
        Then an open Weyl chamber is any of the connected components of
        ${\frak h}-\cup_{\alpha\in\Delta}\Sigma_{\alpha}$.
        Its closure is a closed Weyl chamber.
\end{itemize}
Notice that the closed Weyl chamber is linear isomorphic to the orthant
$$
{\Bbb R}^r_+ =\{(a_1,\ldots,a_r)\mid a_r\ge 0\},
$$
where $r= \dim {\frak h}$.

\subsection{An $\Ad$-invariant polarization in $\mbox{\bigfrak g}-V_0$}

There is a collection of integrable distributions (i.e. a
{\it polarization}) in ${\frak g}$ that arises from the
$\ad$-representation of ${\frak g}$ on itself. It plays an important
role in understanding the T-dual geometry on $\frak g$ and we shall
explain it in some detail.

Let $C$ be a closed Weyl chamber in $\frak g$ and $\Int  C$ be its
interior. Then, from Sec.~3.1.3, one has
$$
{\frak g}-V_0\;=\;\Ad G \cdot\,\Int  C.
$$
Let
$$
{\frak g}={\frak h}\oplus\left(\oplus_{\alpha\in\Delta^{+}}
\Pi_{\alpha}\right)
$$
be as in Sec.~3.1.2 with $C$ lying in ${\frak h}$. Let
$\widehat{\frak h}$, $\widehat\Pi_{\alpha}$ be the distributions along
$\Int  C$ obtained by translating respectively ${\frak h}$, $\Pi_{\alpha}$
over $\Int  C$ using the vector space structure of $\frak g$.
Applying the $\Ad$-action to move them around, one then obtains a
collection of distributions on ${\frak g}-V_0$. Denote the one associated
to $\widehat{\frak h}$ by ${\cal D}_0$ and the one associated to
$\widehat\Pi_{\alpha}$ by ${\cal D}_{\alpha}$. The whole collection
is independent of the choice of Weyl chamber and one has
$$
T_{\ast}\left({\frak g}-V_0\right)\;=\;
  {\cal D}_0\oplus\left(\oplus_{\alpha\in\Delta^{+}}{\cal D}_{\alpha}\right).
$$
This decomposition is orthogonal with respect to both
$\langle\;,\;\rangle$ and $\llangle\;,\;\rrangle$.

\begin{proposition}[Integrability]
The distributions ${\cal D}_0$ and ${\cal D}_{\alpha}$'s
on ${\frak g}-V_0$ are integrable.
\end{proposition}

\noindent{\it Proof:} From the setting, it is clear that ${\cal D}_0$ is
integrable. An integral submanifold of ${\cal D}_0$ is the intersection
of some Cartan subalgebra with ${\frak g}-V_0$. In other words, it is
an open Weyl chamber.

As for ${\cal D}_{\alpha}$, let
$v\,\in\,{\frak g}-V_0$; its stabilizer $\Stab(v)$ under the $\Ad$-action
is the maximal torus $T$ that gives the unique Cartan subalgebra $\frak h$
containing $v$. The $\Ad$-orbit $Q$ through $v$ is diffeomorphic to
$G/T$ and one has
$$
T_{\ast}Q\;=\;\oplus_{\alpha}\left({\cal D}_{\alpha}|_Q\right).
$$
Let
$$
\proj\,:\,G\longrightarrow Q
$$
be the quotient map and $\overline{{\frak h}+\Pi_{\alpha}}$
be the left-invariant distribution on $G$ whose restriction at the
identity is ${\frak h}+\Pi_{\alpha}$. Since ${\frak h}+\Pi_{\alpha}$
is a subalgebra in $\frak g$, $\overline{{\frak h}+\Pi_{\alpha}}$
is integrable. From the fact that
$$
{\cal D}_{\alpha}|_Q\;=\;
  \proj_{\ast}\left(\overline{{\frak h}+\Pi_{\alpha}}\right),
$$
one concludes that
${\cal D}_{\alpha}$ is integrable when restricted to $Q$ and hence it
is integrable in $\frak g$. Its integral submanifolds are the projection
of those for $\overline{{\frak h}+\Pi_{\alpha}}$ in $G$.
This completes the proof.

\hspace{12cm}$\Box$

We shall call either $\{{\cal D}_{\alpha}\}$ or the family of their
integral leaves {\it the polarization
of $\frak g$ indexed by a root system}.

\subsection{Basic properties of the T-dual geometry}

\begin{proposition}
 The dual structures $\llangle\:,\:\rrangle$ and $B$
 on $\frak g$ are both $\Ad$-invariant.
\end{proposition}

\noindent{\it Proof:} After identifying the tangent space at
any $v\in{\frak g}$ with $\frak g$ itself using the vector space
structure, one may write $(\Ad_g)_{\ast}$ for $g\in G$ simply as $\Ad_g$.
With this convention, for $X,\,Y\,\in\,T_v{\frak g}$, one has
\begin{eqnarray*}
\lefteqn{ \llangle \Ad_gX,\Ad_gY\rrangle_{\Ad_gv}\;
          =\;\langle (\Id -\ad_{\Ad_gv})^{-1}\Ad_gX,
                   (\Id -\ad_{\Ad_gv})^{-1}\Ad_gY\rangle }\\
 & & =\;\langle \Ad_g(\Id -\ad_v)^{-1}X,\Ad_g(\Id -\ad_v)^{-1}Y \rangle,
          \mbox{ since } \ad_{\Ad_gv}\;=\;\Ad_g\,\ad_v\,\Ad_g^{-1};\\
 & & =\;\langle (\Id -\ad_v)^{-1}X,(\Id -\ad_v)^{-1}Y \rangle,
        \mbox{ since $\langle\;,\;\rangle$ is $\Ad$-invariant};\\
 & & =\;\llangle X,Y\rrangle_v.
\end{eqnarray*}
And similarly,
\begin{eqnarray*}
\lefteqn{B(\Ad_gX,\Ad_gY)_{\Ad_gv}\;
   =\;\llangle \Ad_gX\,,\,\ad_{\Ad_gv}\,\Ad_gY\rrangle_{\Ad_gv}}\\
 & & =\;\llangle \Ad_gX\,,\,\Ad_g\,\ad_v Y\rrangle_{\Ad_gv}\;
     =\;\llangle X\,,\,\ad_v Y\rrangle_v\\
 & & =\;B(X,Y)_v.
\end{eqnarray*}

\hspace{12cm}$\Box$

\bigskip

\begin{corollary} As a Riemannian submanifold in
$({\frak g},\llangle\:,\:\rrangle)$, every Cartan subalgebra is totally
geodesic.
\end{corollary}

\noindent{\it Proof:} Let $\frak h$ be a Cartan subalgebra in $\frak g$
and $v\,\in\,{\frak h}-V_0$.
Then, for $\varepsilon > 0$ but small enough, $\Ad_{exp(\varepsilon v)}$
is an isometry of $({\frak g},\llangle\:,\:\rrangle)$ whose set of fixed
points is exactly $\frak h$. This shows that $\frak h$ is totally geodesic.

\hspace{12cm}$\Box$

\medskip

Let $v\,\in\,{\frak g}$ then
$\llangle X,X\rrangle|_v\;=\;\langle X,X\rangle_v$ for $X\,\in\,\ker(\ad_v)$.
{}From Sec.~3.2.1, $\ker(\ad_v)$ is ${\cal D}_0|_v$ for
$v\,\in\,{\frak g}-V_0$.
Thus for any tangent vector to $\frak g$ that lies
in ${\cal D}_0$,  its norms
with respect to $\langle\;,\;\rangle$ and $\llangle\;,\;\rrangle$
are the same. Consequently, any path that lies in some
Cartan subalgebra has the same length with respect to either
$\langle\;,\;\rangle$ or $\llangle\;,\;\rrangle$.
Together with the previous corollary then implies that all the
affine lines in $\frak g$ that lie in a Cartan subalgebra
are bi-infinite geodesics with respect to $\llangle\:,\:\rrangle$.
Particularly, all the half lines from the origin are infinite geodesic
rays with respect to $\llangle\:,\:\rrangle$. Thus
the exponential map at the origin with respect to $\llangle\:,\:\rrangle$
$$
\Exp_{O}\,:\,T_{O}{\frak g}\longrightarrow {\frak g}
$$
is well-defined on the whole $T_{O}{\frak g}$.
It actually coincides with the exponential map
with respect to $\langle\;,\;\rangle$.
By Hopf-Rinow theorem \cite{C-E} this shows that

\begin{corollary} $({\frak g},\llangle\:,\:\rrangle)$ is a complete
    metric space.
\end{corollary}

Since $G$ is compact connected, for any $v\,\in\,{\frak g}$,
its stabilizer $\Stab(v)$ under the $\Ad$-action is a connected closed
subgroup in $G$ \cite{He} with
$$
\ker(\ad_v)\;\subset\;T_e \Stab(v).
$$
Since the jump of the dimension of $\ker(\ad_v)$ when varying $v$ is
always even,
$$
\dim  \Stab(v)\;
  \left\{
    \begin{array}{ll}
       =\;r\;(\mbox{ i.e. } \rank G) &\mbox{ if } v\,\in\,{\frak g}-V_0\\
       \ge\;r+2 &\mbox{ if } v\,\in\,V_0.
    \end{array}
  \right.
$$
On the other hand, for any closed Weyl chamber $C$ in a fixed Cartan
subalgebra,
$$
\begin{array}{clll}
  {\frak g}-V_0 & = & \Ad G\,\cdot\,\Int \,C & \mbox{and}\\
  V_0 & = & \Ad G\,\cdot\,\partial C. &
\end{array}
$$
This implies that $V_0$ is a homogeneous variety of codimension $\ge$ 2
and hence the isometric embedding
$$
{\frak g}-V_0 \;\hookrightarrow\;({\frak g},\llangle\;,\;\rrangle)
$$
is distance-preserving. In other words,
$({\frak g},\llangle\;,\;\rrangle)$ is the metric completion
of $({\frak g}-V_0,\llangle\;,\;\rrangle|_{{\frak g}-V_0})$
by a subset of codimension $\ge$ 2 in $\frak g$.
Consequently, the {\it generic part} ${\frak g}-V_0$ itself captures
nearly all the metric properties of the whole
$({\frak g}\,,\,\llangle\:,\:\rrangle)$.

\subsection{Riemannian geometry of the T-dual metric}

Due to the fact that the $\Ad$-action of $G$ on ${\frak g}-V_0$
has stabilizers isomorphic to a maximal torus $T$ of $G$
and that the quotient space is the interior
of a Weyl chamber $C$ which is contractible, one has the following
trivial fibration:
$$
\begin{array}{ccc}
 G/T & \longrightarrow & {\frak g}-V_0\\
    & & \mbox{\hspace{0.5cm}}\downarrow\;\pr\\
    & & \Int  C\;.
\end{array}
$$
Also recall the common $\Ad$-invariant orthogonal decomposition
$$
T_{\ast}\left({\frak g}-V_0\right)\;=\;
  {\cal D}_0\oplus\left(\oplus_{\alpha\in\Delta^{+}}{\cal
D}_{\alpha}\right)
$$
with respect to both $\langle\;,\;\rangle$ and $\llangle\;,\;\rrangle$.
The following proposition shows that the T-dual metric on $\frak g$
is a {\it polarized conformal deformation} of the flat Killing
metric and the generic part is a {\it polarized warped-product}
of the flat cone $\Int  C$ with $G/T$.

\begin{proposition} Let
$$
\langle\;,\;\rangle\;=\;ds_0^2\:+\:\sum_{\alpha\in\Delta^{+}}\,ds_{\alpha}^2
$$
with respect to the above decomposition. Then, for any
$v\,\in\,{\frak g}-V_0$,
$$
\llangle\;,\;\rrangle_v\;=\;
  ds_0^2|_v\:+\:\sum_{\alpha\in\Delta^{+}}\,
    \frac{1}{1\,-\,\alpha(\bar{v})^2}\,ds_{\alpha}^2|_v,
$$
where $\bar{v}\;=\;\pr(v)\;\in\;\Int  C$.
\end{proposition}

\noindent{\it Proof:} Since ${\cal D}_0$, ${\cal D}_{\alpha}$'s
are invariant under the $\Ad$-action, the decomposition of
$\langle\;,\;\rangle$ is also invariant under the $\Ad$-action.
Without loss of generality, we may assume that the open Weyl chamber
$\Int  C$ is embedded in ${\frak g}-V_0$ and contains $v$.
Now for $X\,\in\,{\cal D}_{\alpha}$, $Y\,\in\,{\cal D}_{\beta}$
($\alpha$, $\beta$ could be 0), one has
\begin{eqnarray*}
\lefteqn{\llangle X,Y\rrangle\;=\;
  \langle (\Id  - \ad_v)^{-1}X, (\Id  - \ad_v)^{-1}Y\rangle}\\
  & & =\;\langle X, (\Id  + \ad_v)^{-1}(\Id  - \ad_v)^{-1}Y\rangle\\
  & & =\;\langle X, \left( \Id  - (\ad_v)^2\right)^{-1}Y\rangle.
\end{eqnarray*}
Notice that $(\ad_v)^2$ is symmetric with respect to $\langle\;,\;\rangle$.
The eigenspace decomposition of $T_v{\frak g}$ for $(\ad_v)^2$ coincides with
$\left({\cal D}_0\oplus
\left(\oplus_{\alpha\in\Delta^{+}}{\cal D}_{\alpha}\right)\right)|_v$.
For $Z\,\in\,{\cal D}_{\alpha}$,
$$
(\ad_v)^2(Z)\;=\;\alpha(v)^2\,Z.
$$
Thus
\begin{eqnarray*}
\lefteqn{\llangle X,Y\rrangle\;=\;\langle X, (1 - \beta(v)^2)^{-1}Y\rangle}\\
  & & =\;\frac{\delta_{\alpha\beta}}{1\,-\,\alpha(\bar{v})^2}
         \langle X,Y\rangle.
\end{eqnarray*}
This concludes the proof.

\hspace{12cm}$\Box$

Notice that $\alpha(\bar{v})$ is purely imaginary, thus
$$
0\;<\;\frac{1}{1\,-\,\alpha(\bar{v})^2}\;\le\;1.
$$
Since ${\frak g}-V_0$ is open and dense in $\frak g$, we have
\begin{corollary}
 For any $X\,\in\,T_{\ast}{\frak g}$, $\llangle X,X\rrangle\;\le\;
                                \langle X,X\rangle$.
\end{corollary}

As $v$ approaches the characteristic variety $V_0$, some of the
$\alpha(\bar{v})$'s gets closer and closer to 0. In the limit,
their corresponding ${\cal D}_{\alpha}$'s are absorbed into
the undistorted flat directions at the limit point in $V_0$.

\begin{figure}
 \centerline{\figbox{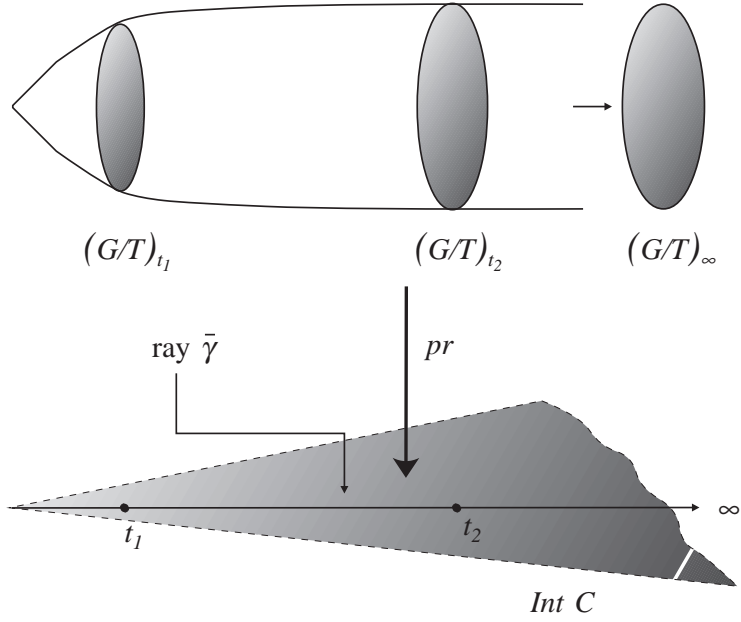}{11cm}}
 \caption[dual-geom]{Asymptotic stability of the T-dual
        Riemannian geometry. With a neighborhood of $V_0$ deleted,
        the rest of the metric space looks like a cone
        on a sufficiently large scale.}
  \protect\label{dual-geom}
\end{figure}

The explicit expression in the set of the polarized conformal factor
$\left\{\frac{1}{1\,-\,\alpha(\bar{v})^2}\right\}$ together with the
polarized warped-product structure on ${\frak g}-V_0$ gives a clear
picture of what the T-dual $\frak g$ looks like as a Riemannian
manifold ({\sc Figure~\ref{dual-geom}}). We summarize them partially as
\begin{proposition}[Asymptotic stability]
  Let $\bar{\gamma}(t)$ be a ray in $\Int  C$ from the origin
  parameterized by arc-length. Let $(G/T)_t$ be the fiber over
  $\bar{\gamma}(t)$ in the polarized warped-product.
  Then $(G/T)_t$ is a polarized conformal deformation of $(G/T)_1$
  with polarization $\{\cal D_{\alpha}\}$ and family of factors
  $$
   \left\{\frac{t^2\,\left(1\,-\,\alpha(\bar{\gamma}(1))^2\right)}
            {1\,-\,t^2\,\alpha(\bar{\gamma}(1))^2}\right\}_{\alpha}.
  $$
  Consequently, $(G/T)_{\infty}$ is a polarized conformal deformation
  of $(G/T)_1$ with factors
  $\left\{-\,\frac{1}{\alpha(\bar{\gamma}(1))^2}\right\}_{\alpha}$
  and hence is compact.
  With a neighborhood of $V_0$ deleted from $\frak g$, the rest is
  quasi-isometric \cite{Gr} to the base cone with a neighborhood of boundary
  deleted.
\end{proposition}
\medskip

\noindent{\it Proof}. All this follows from the fact that
the collection of Riemannian manifolds $\{(G/T)_t\}$ forms a
radial infinite cone at the origin with base $(G/T)_1$. Using the
radial projection from $(G/T)_t$ to $(G/T)_1$ and the invariance of
${\cal D}_{\alpha}$ under radial scaling maps, one immediately
justifies all the claims. This concludes the proof.

\hspace{12cm}$\Box$

\noindent{\it Remark}. When applied to the special pair,
$SU(2)$ and ${\frak su}(2)$,
$$
V_0\,=\,\mbox{the origin},\;\Int C\,=\,\mbox{ a half line }L_+,\;
\mbox{ and }\;G/T\,=\,S^2.
$$
The proposition says that ${\frak su}(2)-\{\mbox{origin}\}$ with the T-dual
metric is a warped-product of $L_+$ with $S^2$ with factor
$\frac{t^2}{1+4t^2}$, whose limit is $\frac{1}{4}$ as
$t\rightarrow\infty$. One can check that this coincides with the known
results from the literature.

\subsection{The B-field}

Recall that, with respect to the $\Ad$-invariant metric
$\langle\:,\:\rangle$, $\frak g$ has a Poisson structure given by
a closed 2-form $\zeta$ with
$$
\zeta_v(X,Y)\;=\;\langle X,\ad_v\,Y\rangle,
$$
for $X$, $Y$ in $T_v{\frak g}$. Its symplectic leaves are the
$\Ad$-orbits. Analogous to the T-dual metric, $B$ can be written as
a polarized conformal deformation of $\zeta$. Explicitly, assuming that
$v$ is in ${\frak g}-V_0$, let $X,\,Y\,\in T_v{\frak g}$ and decompose
$$
 X\,=\,\sum_{\alpha}X_{\alpha},\;Y\,=\,\sum_{\alpha}Y_{\alpha},
$$
where $X_{\alpha},\,Y_{\alpha}\,\in {\cal D}_{\alpha}$ ($\alpha$ could be
zero here).
Straightforward computation then gives
$$
B(X,Y)\;=\;\sum_{\alpha}\,\frac{1}{1\,-\,\alpha(v)^2}\,
 \zeta(X_{\alpha},Y_{\alpha}),
$$
which, for the special case of $SU(2)$ - ${\frak su}(2)$ pair, again gives the
known $B$-field. However,
direct checking shows that, after this polarized distortion,
$B$ is no longer closed for general simple Lie algebras;
nor do the $\Ad$-orbits remain symplectic in general. It's not clear
to us at the moment what kind of geometry this $B$-field provides
on $\frak g$ in general.


\newpage
\begin{thebibliography}{MMMMMMM}

\bibitem[A-AG-B-L]{A-AG-B-L}
E.~Alvarez, L.~Alvarez-Gaum\'{e}, J.~Barb\'{o}n and Y.~Lozano,
``Some global aspects of duality in string theory'', Nucl. Phys. B415:71-100
(1994), {\tt hep-th/9309039}.

\bibitem[A-AG-L1]{A-AG-L1}
E. Alvarez, L. Alvarez-Gaum\'e and Y. Lozano,
``On Non-abelian Duality'', Nucl. Phys. B424:155-183 (1994),
{\tt hep-th/9403155}.

\bibitem[A-AG-L2]{A-AG-L2}
E.  Alvarez, L.  Alvarez-Gaum\'e and Y.  Lozano, ``A canonical approach
to duality transformations''.  Phys. Lett. B336:183-189 (1994), {\tt
hep-th/9406206}.


\bibitem[A-G-M]{A-G-M} P.S. Aspinwall, B.R. Greene and D. Morrison,
  ``Space-time topology change and stringy geometry'',
  J. Math. Phys. 35 (1994), pp. 5321 - 5337.



\bibitem[A-M]{A-M} R. Abraham and J.E. Marsden,
  {\it Foundations of mechanics}, 2nd ed., Benjamin/Cummings Publ.,
  1978.




\bibitem[Ar]{Ar} V.I. Arnold,
  {\it Mathematical methods of classical mechanics}, 2nd ed., GTM 60,
  Springer-Verlag, 1989.

\bibitem[B1]{B1}
T.H. Buscher, ``A symmetry of the string background field equations'',
Phys. Lett. 194B: 59 (1987) .

\bibitem[B2]{B2} T.H. Buscher,
``Path integral derivation of quantum duality in nonlinear sigma models'',
Phys. Lett. 201B: 466 (1988).




\bibitem[C-E]{C-E} J. Cheeger and D. Ebin,
  {\it Comparison theorems in Riemannian geometry}, North-Holland, 1975.



\bibitem[C-Z]{C-Z} T. Curtright and C. Zachos,
  ``Currents, charges, and canonical structure of pseudodual chiral
  models'', Phys. Rev. D49 (1994), pp. 5408 - 5421.

\bibitem[dlO-Q]{dlO-Q}
X.  de la Ossa and F.  Quevedo, ``Duality symmetries from nonabelian
isometries in string theory'', Nucl.  Phys.  B403 (1993) 377, {\tt
hep-th/9210021}.


\bibitem[E-G-R-S-V]{E-G-R-S-V}
S.  Elitzur, A.  Giveon, E.  Rabinovici, A.  Schwimmer, G.  Veneziano,
``Remarks on nonabelian duality'',.  Nucl. Phys. B435:147-171,1995, {\tt
hep-th/9409011}.


\bibitem[F-J]{F-J}
B.E.  Fridling and A.  Jevicki, ``Dual representations and ultraviolet
divergences in nonlinear sigma models'', Phys.  Lett.  134B (1984) 70.



\bibitem[G-R-V]{G-R-V}
M. Gasperini, R. Ricci and G. Veneziano,
``A problem with nonabelian duality?''
Phys. Lett. B319 (1993) 43, {\tt hep-th/9308112}.


\bibitem[G-K]{G-K}
A.  Giveon and E.  Kiritsis, ``Axial vector duality as a gauge symmetry and
topology change in string theory'', Nucl.  Phys.  B411 (1994) 487, {\tt
hep-th/9303016}.


\bibitem[G-P-R]{G-P-R} A. Giveon, M. Porrati and E. Rabinovici,
  {\it Target space duality in string theory},
  Phys. Reports 244(1994), pp. 77 - 202, {\tt hep-th/9401139}.

\bibitem[G-R-V]{G-R-V}
A.  Giveon, E.  Rabinovici and G.  Veneziano, ``Duality in string
background space'', Nucl.  Phys.  B322 (1989) 167.



\bibitem[G-R1]{G-R1}
A.  Giveon and M.  Ro\u{c}ek, ``Generalized duality in curved string
backgrounds'', Nucl.  Phys.  B380 (1992) 128, {\tt hep-th/9112070}.

\bibitem[G-R2]{G-R2}
A. Giveon and M. Ro\u{c}ek, ``On Nonabelian Duality'',
Nucl. Phys. B421:173, (1994), {\tt hep-th/9308154}.


\bibitem[G-R3]{G-R3} A. Giveon and M. Ro\u{c}ek,
  ``Introduction to duality'',
  {\tt hep-th/9406178}.

\bibitem[Gr]{Gr} M. Gromov,
  {\it  Structures m\'etrique pour les vari\'et\'es riemanniennes},
  r\'edig\'e par J.Lafontaine et P.Pansu, Text Math. no. 1,
  Cedic/Fernand--Nathan, Paris, 1980.


\bibitem[He]{He} S. Helgason,
  {\it Differential geometry, Lie groups and symmetric spaces},
  Academic Press, 1978.

\bibitem[K-S1]{K-S1} C.~Klim\v{c}\'{\i}k and P.~\v{S}evera,
``Strings in spacetime cotangent bundle and T-duality'',
{\tt hep-th/9411003}.

\bibitem[K-S2]{K-S2} C.~Klim\v{c}\'{\i}k and P.~\v{S}evera,
``Dual and non-abelina duality and the Drinfeld double'',
{\tt hep-th/9502122}.

\bibitem[Lo]{Lo} Y. Lozano,
  {\it Non-abelian duality and canonical transformation},
{\tt   hep-th/9503045}.


\bibitem[M-V]{M-V}
K.A.  Meissner and G.  Veneziano,
``Symmetries of cosmological superstring vacua'',
Phys.  Lett.  B267 (1991) 33.



\bibitem[Mi]{Mi} J. Mickelsson,
  {\it Current algebras and groups}, Plenum Press, 1989.


\bibitem[P-S]{P-S} A. Pressley and G. Segal,
  {\it Loop groups}, Clarendon Press, Oxford, 1990.

\bibitem[R-V]{R-V}
M.  Ro\u{c}ek and E.  Verlinde, ``Duality, quotients, and currents'', Nucl.
Phys.  B373 (1992) 630, {\tt hep-th/9110053}.



\bibitem[Sa]{Sa} H. Samelson,
  {\it Notes on Lie algebras}, Springer-Verlag, 1990.



\end{thebibliography}
\end{document}